%% file: CMBpol_gbase.tex
\documentstyle[psfig,subfigure]{mn}

\newcommand{\rgl}{\rangle}
\newcommand{\lgl}{\langle}
\newcommand{\A}{\mbox{\boldmath $A$}}
\newcommand{\Xib}{\mbox{\boldmath $\Xi$}}
\newcommand{\Fc}{{\cal F}}
\newcommand{\QU}{QUaD}

\begin{document}
\input{epsf.tex}
\title[Optimization of a ground-based CMB polarization experiment]
{Scientific optimization of a ground-based CMB polarization experiment}
\input{author_list.tex}
\maketitle

\begin{abstract}
We investigate the science goals achievable with the upcoming
generation of ground-based Cosmic Microwave Background
polarization experiments, focusing on one particular
experiment, \QU, a proposed bolometric polarimeter
operating from the South Pole. We calculate the optimal sky coverage
for this experiment including
the effects of foregrounds and gravitational lensing. We find that an E-mode measurement will be sample-limited, while
a B-mode measurement will be detector-noise-limited. We conclude that
a 300 deg$^2$ survey is an optimal compromise for a two-year experiment to measure 
both E and B-modes, and that a 
ground-based polarization experiment can make an important
contribution to B-mode surveys. \QU{} can make a
high significance measurement of the acoustic peaks in the E-mode
spectrum, over a multipole range of  $25<\ell <2500$, and will be
able to detect the gravitational lensing signal in the B-mode
spectrum. Such an experiment could also directly detect the
gravitational wave component of the B-mode spectrum if the
amplitude of the signal is close to current upper limits. We also
investigate how \QU{} can improve
constraints on the cosmological parameters. We estimate that combining two years of \QU{} data with the
four-year {\it WMAP} data
can improve constraints on $ \Omega_b h^2 $, $ \Omega_m
h^2 $, $ h $, $r$ and $ n_s $ by a factor of two.
If the foreground contamination can be reduced, the measurement of $r$
can be improved by up to a factor of six over that
obtainable from {\it WMAP} alone. These
improved accuracies will place strong constraints on the
potential of the inflaton field.
\end{abstract}

\begin{keywords}
cosmic microwave background --
cosmological parameters  -- \\ methods:observational
-- polarization -- techniques:polarmetric
\end{keywords}

\clearpage 

\section{Introduction}
\label{sec:intro} The Cosmic Microwave Background (CMB) has proven
to be a powerful cosmological probe. Successive generations of
experiments have provided a stringent test for the standard Big
Bang paradigm and increasingly sensitive measurements of the
temperature power anisotropies have led to tight constraints on
many of the fundamental cosmological parameters. However, as well
as fluctuations in the CMB temperature field, there are also
anisotropies in the linear polarization of the CMB. These
polarization fluctuations have recently been detected by the DASI
experiment \cite{DASI} and the correlation between the temperature
and polarization has been measured by the Wilkinson Microwave Anisotropy Probe ({\it WMAP}) satellite
\cite{WMAP:TE}. However, to make full use of the CMB, higher
sensitivity high resolution polarized measurements are needed.
This is the challenge facing the next generation of CMB
experiments.

While it is desirable to observe the CMB temperature field from
space, to remove atmospheric noise, this is not as important for
polarization experiments since the atmospheric emission is not
expected to be linearly polarized \cite{BK}.  Therefore, by
integrating deeply on relatively small patches of sky (Jaffe, Kamionkowski \& Wang, 2000) it is
possible to make a measurement of the polarization anisotropies
with a comparable signal-to-noise ratio to a satellite experiment
on all but the largest angular scales. 

The survey design for a ground-based experiment will depend upon
the specific science goals of the experiment.  In this paper we
investigate observing strategies and sky coverage for the
forthcoming generation of ground-based CMB polarization experiments,
taking into account foreground issues. We will also show how
ground-based polarization measurements can help to tighten
constraints on the cosmological model. 

To be concrete, we focus on one particular experiment,
\QU{} (QUEST: Q and U Extra-galactic Sub-millimetre Telescope,
and DASI: Degree Angular Scale Interferometer). This is a proposal 
to install QUEST\footnote{http://www.astro.cf.ac.uk/groups/instrumentation/projects/},
a high-resolution bolometric array polarimeter, on the
azimuth-elevation mount of the DASI\footnote{http://astro.uchicago.edu/dasi/} instrument.
The experiment plans to begin observing from the South Pole in $2005$ \cite{SEC}.

The remainder of the paper is set out as follows. In Section
\ref{sec:polrev} we briefly review the physics of the CMB
polarization. In Section \ref{sec:formal} we present the formalism
used in the investigation and in Section \ref{sec:fg} we show how
we have included the effects of foregrounds. Our cosmological
model and definitions are presented in Section \ref{sec:cosmod}. In Section \ref{sec:areaopt} we present our
results for the survey design, and in Section \ref{sec:maps} simulate
polarization maps. The expected accuracies and multipole coverage
of the power spectra are presented in Section \ref{sec:spec}, and in Section
\ref{sec:param} we present the expected parameters constraints for
the \QU{} experiment. Our findings are summarized in Section \ref{sec:discuss}. We 
also include an appendix in which we discuss the sensitivity definitions
used in our calculations.
We begin with a brief review of CMB polarization.

\section{Review of CMB polarization}
\label{sec:polrev}
Detailed reviews of the CMB polarization are given by Zaldarriaga
(2003) and Hu \& White (1997). In this Section we give a brief
overview of how the polarization field is generated and how it is
parameterized.

\subsection{Parameterization of the polarization field}

Typically, a linearly polarized source is quantified by the Q and U Stokes parameters, which
give the differences in intensity between orthogonal polarization states. These quantities are convenient to 
measure experimentally, but are not invariant under a rotation of the co-ordinate system
and so are difficult to compare to theoretical models. It is useful to define the quantities $Q(\theta,\phi) \pm iU(\theta,\phi)$
which under rotation acts as a spin 2 quantity. By operating on $Q \pm iU$ using spin raising and lowering operaters it is possible to obtain two rotationally invariant spin 0 fields,
$E(\theta,\phi)$ and $B(\theta,\phi)$ (Zaldarriaga \& Seljak, 1997). This represents the decompostion of the polarization field into different parity states, the E field 
is unchanged by a parity transformation, but the B field changes sign. These E and B fields can then be compared directly to theoretical predictions\footnote{An equivalent formalism is given by
Kamionkowski, Kosowsky \& Stebbins (1997) in which the polarization field is
expanded in terms of tensor spherical harmonics instead of spin 2
harmonics.}. 

The temperature field is usually
expanded in terms of scalar spherical harmonics, $  Y_{\ell m}(\theta,\phi) $:
\begin{equation}
\frac{\Delta T(\theta,\phi)}{T_o}= \sum_{\ell m}T_{\ell
m}Y_{\ell m}(\theta,\phi),
\end{equation}
where $\Delta T$ is the deviation of the temperature field from
its average value $T_o$. Similarly, the quantities $Q \pm iU$ can be expanded in terms of spin 2
spherical harmonics, $_{\mp 2}Y_{\ell m} $:
\begin{equation}
Q(\theta,\phi) \pm iU(\theta,\phi)=\sum_{\ell m}
\left(
E_{\ell m} \mp i B_{\ell m}
\right)_{\mp 2}Y_{\ell m}(\theta,\phi).
\end{equation}

The two point statistics of the CMB can be completely
described in terms of the covariances of the multipole moments, $
T_{\ell m}, E_{\ell m}$ and $B_{\ell m} $:
\begin{equation}
\label{eq:multipole}
\begin{array}{cc}

\left< T_{\ell m}^* T_{\ell' m'} \right>
=C_{\ell}^{TT}\delta_{\ell \ell'}\delta_{m m'}
&
\left< E_{\ell m}^*E_{\ell' m'} \right>
=C_{\ell}^{EE}\delta_{\ell \ell'}\delta_{m m'}
\\
\left< B_{\ell m}^*B_{\ell' m'} \right>
=C_{\ell}^{BB}\delta_{\ell \ell'}\delta_{m m'}
&
\left< T_{\ell m}^*E_{\ell' m'} \right>
=C_{\ell}^{TE}\delta_{\ell \ell'}\delta_{m m'}
\\
\left< T_{\ell m}^*B_{\ell' m'} \right>
=C_{\ell}^{TB}\delta_{\ell \ell'}\delta_{m m'}
&
\left< E_{\ell m}^*B_{\ell' m'} \right>
=C_{\ell}^{EB}\delta_{\ell \ell'}\delta_{m m'}.

\label{eq:spec}
\end{array}
\end{equation}
As the B field has opposite parity to the T and E fields
the TB and EB correlations are zero if we can assume
that parity is conserved. If the CMB is a Gaussian
random field, as predicted if the metric fluctuations are
generated by zero-point fluctuations during inflation, the statistical
properties of the CMB
temperature and polarization fields are completely defined by the
four power spectra, $ C_{\ell}^{TT}, C_{\ell}^{EE}, C_{\ell}^{BB}
$ and $ C_{\ell}^{TE} $. However, as we shall discuss in Section
2.3, gravitational lensing by large-scale structure along the line
of sight will distort the pattern of fluctuations, and will induce
non-Gaussianity.

\subsection{Polarization signal generated during recombination}
The CMB polarization signal primarily arises from the Thomson
scattering of the CMB photons during recombination. Polarization
can only be generated if the radiation field contains a local
quadrupole.  Density perturbations will produce a velocity
gradient in the primordial plasma so that photons approaching an
electron from different directions will be Doppler shifted by
different amounts. This produces local quadrupoles in the
radiation field. Before recombination, the high electron density
means that the mean free path of the photons is too small to
produce a quadrupole; however, after the recombination the
electron density is too low for significant Thomson scattering to
occur.  The polarization can only be produced during a
short period around recombination, so the amplitude of the
polarization is very low.

The mechanism by which these scalar perturbations are produced in
the polarization field is therefore subtly different to the way in
which the temperature perturbations are produced. A measurement of
the polarization power spectra will not only provide a
consistency check of the cosmological model, but will also yield
new information on processes occurring in the early universe. Much
of this information is contained in the TE and EE acoustic peaks
at high $ \ell $ which can be measured with high signal to noise
with a ground-based experiment.

The inflationary model also predicts a stochastic background of
gravitational waves (GW) which will also result in a quadrupole.
The decomposition of the polarization field into the E and B modes
can be used to separate the GW (tensor) contribution from the
density perturbation (scalar) contribution.  The E-modes can be
produced by both scalar and tensor perturbations, but the B-modes
produced at last scattering can only be generated by tensor
perturbations. This means that a measurement of the B-mode
spectrum would give new information about inflationary parameters.
In particular, the amplitude of the tensor spectrum is directly
related to the energy scale of inflation. These parameters
can not be well constrained from the TT and EE spectrum as it
difficult to separate the tensor and scalar contributions to these
measurements.  The GW B-mode signal peaks around scales of about
 $ \ell=100 $ and so in principle is detectable from the ground.

\subsection{Polarization signal generated after recombination}
\begin{figure}
\psfig{file=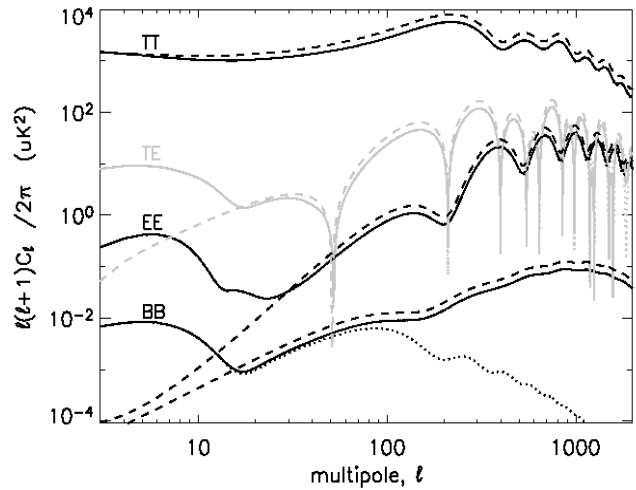,width=8.92cm,height=7.0cm}
\caption{CMB temperature and polarization power spectra. The
dashed lines are for a model with no re-ionization while the
dotted lines are for a model with no gravitational lensing. The
solid lines include the effects of  both gravitational lensing and
re-ionization.  Model parameters are given in Section \ref{sec:cosmod}.}
\label{fig:secondary}
\end{figure}

The polarization spectra generated at recombination will be
altered mainly by two processes before they can be detected:
re-ionization and weak gravitational lensing (GL).  The effect of
re-ionization is to increase the polarization signal on large
scales ($ \ell \le 20$). Ground-based experiments are unlikely to
be able to measure the polarization on such large angular scales
and so will not be sensitive to the effects of re-ionization.
However, weak lensing affects the signal on small angular scales.
CMB photons are deflected by the gravitational potential of large
scale structure.  For the TT and EE spectra, this effect results
in a  smearing of the acoustic peaks on small angular scales,
although the change to the spectra is very small, as shown in Fig.
\ref{fig:secondary}. However, lensing will also convert E-mode
polarization into B-modes.  This means that there will be a scalar
contribution to the B-mode spectrum due to lensing.  Therefore,
the B-mode spectrum will be contaminated by a GL contribution, the
spectrum of which must be measured precisely so that it can be
removed \cite{KS}. As the lensing signal peaks at small angular
scales, ground-based experiments are well-suited to this task.

The lensing signal itself also contains useful information about
large-scale structure. This can be used to constrain other
cosmological parameters such as the neutrino mass \cite{neutrino},
since this will add to the mass-energy of the universe, altering
its expansion history, and suppressing small scale power in the matter power
spectrum due to free streaming. The lensing signal will also make
the CMB sensitive to the equation of state of the universe,
parameterized by $w=p/\rho$, as again this will affect the
expansion history.

Fig. \ref{fig:secondary} shows the temperature and polarization
power spectra, generated by the Boltzmann and Einstein solver
{\small CMBFAST} (version 4.2; Seljak \& Zaldarriaga (1996)\footnote{http://www.cmbfast.org/},
decomposed into temperature-temperature
(TT) power, temperature-E-mode (TE) cross-power, E-mode-E-mode
(EE) power, and B-mode-B-mode (BB) power. We plot spectra without
gravitational lensing (dotted lines) and without re-ionization
(dashed lines) and with both included (solid lines). The main aim
of this paper is to determine the best survey design for the measurement of the polarization
spectra.  In the following section we present our formalism for
this procedure, based on the Fisher Information matrix.

\section{Formalism}
\label{sec:formal}
\subsection{Fisher Information matrix}
For a model dependent on a set of parameters,
\mbox{\boldmath{$\alpha$}}, the probability of a particular
parameter set, given a set of experimental data points, $\bf{d}$,
is expressed by the likelihood function,
$L($\mbox{\boldmath{$\alpha$}}$\mid\!\!\bf{d})$, the probability
of the parameters given the data. By exploring the parameter space
to maximize $ L $ we may determine the parameter values within
certain error limits.  The minimum possible variance with which a
parameter can be measured can be estimated from the Fisher
information matrix \cite{Andy}, defined as:
\begin{equation}
\label{eq_fish}
\Fc_{ij}=\left<\frac{\partial^2{\cal L}}{
\partial \alpha_{i}\partial \alpha_{j}}\right>,
\end{equation}
where $ {\cal L}=\! -\ln L $ and the derivatives are evaluated at
the maximum likelihood values of the parameters.  The inverse of
the Fisher matrix gives the parameter covariance matrix, ${\cal
C}_{ij}$, for the theoretical parameters:
\begin{equation}
{\cal C}_{ij} \equiv \lgl \Delta \alpha_i \Delta \alpha_j \rgl
=\Fc_{ij}^{-1},
\end{equation}
where $\Delta \alpha_i$ is the deviation of the parameter from its
maximum likelihood value. The diagonal of the inverse Fisher
matrix yields the marginalized 1-$\sigma$ error on the parameters.
Taking the inverse of the diagonal of the Fisher matrix,
\begin{equation}
(\Delta \alpha_{i})^2=1/\Fc_{ii},
\end{equation}
yields the conditional error on the parameters. In general
 \begin{equation}
    [\Fc^{-1}]_{ii} \ge 1/\Fc_{ii},
 \end{equation}
 where the equality holds only for uncorrelated parameters. The
 Fisher matrix then provides a theoretical upper bound on the
 accuracy of a measurement of a given parameter for a given experiment.

\subsection{Application of Fisher matrix to CMB experiments}

For a CMB experiment, the data are the measurements of the four
CMB power spectra and the parameters are the cosmological
parameters. For the measurement of a single power spectrum, $
C_{\ell} $, the Fisher matrix is given by:
\begin{equation}
\label{eq:fishs} \Fc_{ij}=\sum\limits_{\ell} \frac{1}{(\Delta
C_{\ell})^2} \frac{\partial C_{\ell}}{\partial \alpha_i}
\frac{\partial C_{\ell}}{\partial \alpha_j},
\end{equation}
where
 \begin{eqnarray}
    (\Delta C_{\ell})^2 = \frac{2}{(2 \ell +1) f_{sky} \Delta \ell} (C_\ell
    + N_\ell)^2 \nonumber
  \end{eqnarray}
 is the error in the measurement of the
power spectrum in a band centred on multipole $ \ell $, and
$N_\ell$ is a noise term. The survey area is given by $f_{sky}$.
The summation is over pass-bands of width $\Delta \ell$.

 For a measurement of all four power
spectra this generalizes to \cite{ZS}:
\begin{equation}
\label{eq:fishc} \Fc_{ij}=\sum\limits_{\ell}\sum\limits_{XY}
\frac{\partial C_{\ell}^{X}}{\partial \alpha_i} \left[\,
\Xib_{\ell}\right]_{XY}^{-1} \frac{\partial C_{\ell}^{Y}}{\partial
\alpha_j},
\end{equation}
where $X$ and $Y$ are either TT, EE, TE or BB and $  \Xi_{XY}
\equiv {\rm Cov}(C^X_\ell C^Y_\ell) $ is the power spectra
covariance matrix:
\begin{equation}
\label{eq:PScov} {\Xib}_{\ell}= \left(\begin{array}{cccc}
{\Xi}_{\ell}^{TT,TT} & {\Xi}_{\ell}^{TT,EE} & {\Xi}_{\ell}^{TT,TE} & 0 \\
{\Xi}_{\ell}^{TT,EE} & {\Xi}_{\ell}^{EE,EE} & {\Xi}_{\ell}^{EE,TE} & 0 \\
{\Xi}_{\ell}^{TT,TE} & {\Xi}_{\ell}^{EE,TE} & {\Xi}_{\ell}^{TE,TE} & 0 \\
0 & 0 & 0 & {\Xi}_{\ell}^{BB,BB}
\end{array}\right).
\end{equation}
The terms in the power spectra covariance matrix are given by:
 \begin{eqnarray}
 \label{eq:XYXY}
 \Xi_{\ell}^{xy,x'y'}&=&\frac{1}{(2\ell+1)f_{sky}\Delta \ell} \nonumber \\
    & \times& \big[(C_{\ell}^{xy'}+N_{\ell}^{xy'})
        (C_{\ell}^{yx'}+N_{\ell}^{yx'})\nonumber \\
    & & + (C_{\ell}^{xx'}+N_{\ell}^{xx'})
        (C_{\ell}^{yy'}+N_{\ell}^{yy'})\big],
 \end{eqnarray}
 where $(x,y)=(T,E,B)$.
 The noise covariance is given by $N_{\ell}^{xy}$. In the case of
no foregrounds this is given by:
\begin{equation}
N_{\ell}^{xy}=w_x^{-1}|{\cal B}_{\ell}^x|^{-2}\delta_{xy},
\end{equation}
where, $ w_x^{-1}=\Omega_{pix}^x(\sigma_{pix}^x)^2$, for an
experiment with solid angle per pixel, $ \Omega_{pix} $, and the
noise per pixel, $ \sigma_{pix} $. The pixel noise depends on
survey design and instrument parameters.  For an experiment
covering an area $ \Theta^2 $ for an integration time $ t_{obs} $,
with $ N_{\rm PSB} $ detectors, a solid angle per pixel $ \Omega_{pix} $ and a
sensitivity\footnote{The definition of sensitivity for a
polarization experiment is discussed in Appendix (A).}, NET, the
pixel noise is:
\begin{equation} \label{eq:pnoise}
\sigma_{pix}^2=\frac{{\rm NET}^2\Theta^2}{t_{obs}N_{\rm
PSB}\Omega_{pix}}.
\end{equation}
In Section \ref{sec:fg} we discuss how the noise terms may be
extended to include foregrounds. We assume that the pixel size used in the map wil be the 
same as the beam size of the telescope. The spherical harmonic transform
of the beam is given by ${\cal B}_{\ell} $. Here we assume that
the beam is a Gaussian,
\begin{equation}
{\cal B}_{\ell}=\exp\left(-\ell(\ell+1)\sigma_{\cal B}^2/2\right),
\end{equation}
with $\sigma_{\cal B}=\theta_{\cal B}/\sqrt{8\ln 2}$ where
$\theta_{\cal B}$ is the full width half maximum beam size.
The pixel size can then be approximated by $\Omega_{pix}=\theta_{\cal B}^2$.

The minimum resolution of the power  spectra, $ \Delta \ell $,
depends on the area of sky covered, $ \Delta \ell=\pi/\Theta $.
This will therefore also give the minimum $ \ell $ at which the
power spectra can be measured as discussed further in Section \ref{sec:QU}.
If a resolution smaller than this
is used, the different $ \ell $ modes will become correlated and
equation (\ref{eq:multipole}) will no longer apply \cite{Hobson}.
We calculate the maximum $ \ell $ value from the FWHM beam size, $
\ell_{max}=\pi/\theta $.  In reality, multipoles higher than this could
be measured if the beam profiles can be accurately determined.

The Fisher matrix also provides a simple way to calculate the results
obtainable by combining a number of observations from different CMB experiments.
In the simplest case, in which $N_{exp}$ experiments observe different patches of sky,
the combined Fisher matrix, $ \Fc^{C} $, is the sum of the individual Fisher matrices, 
 $\Fc^{e}$ \cite{HuFish}:
\begin{equation}
\Fc_{ij}^C=\sum\limits_{e=1}\limits^{N_{exp}}\Fc_{ij}^e.
\end{equation}
If any of the patches of sky overlap, each overlapping region is 
considered as a separate patch. In these patches the combined
noise covariance, $ N_{\ell} $, of the overlapping experiments should 
be used to calculate the terms in the power spectra covariance matrix (equation (\ref{eq:XYXY})).
This is discussed further in the next section
where we consider how to optimally combine multi-frequency data.

This completes the formal machinery we will require for our
analysis. Note that we have ignored the effects of windowing and
mode-mixing due to limited sky coverage (e.g. Bunn 2002), and
non-Gaussianity and mode-coupling induced by gravitational lensing
(e.g. Guzik, Seljak \& Zaldarriaga 1999). The former effects modes
by convolving them with the survey window function and mixing $E$
and $B$ modes. This will mainly effect the B-modes, where the
signal-to-noise is poor, and will slightly increase our
uncertainties. Non-Gaussianity induced by gravitational lensing
will also correlate modes and will give rise to higher-order
correlations, which will also lead to a slight increase in our
uncertainties.

So far we have also ignored the effects of foreground
contamination, and it is to this we now turn.

\section{Foregrounds}
\label{sec:fg}
\input{foreground.tex}

\section{Cosmological model} \label{sec:cosmod}
In order to calculate the terms in the power spectrum covariance
matrix and the power spectrum derivative we require a model from
which to calculate the CMB power spectra.  This model is defined
by two sets of parameters: the inflationary parameters which
parameterize the initial perturbations causing the fluctuations in
the CMB, and the cosmological parameters which determine how these
initial perturbations are propagated into the observed CMB power
spectra. Given a set of parameters, the CMB power spectra can then
be calculated using a Boltzmann and Einstein solver. For this work
we have used a slightly modified version of CMBFAST $ v4.2 $.

The initial scalar perturbations are parameterized by:
\begin{equation}
\Delta_{\cal R}^2(k)=\Delta_{\cal R}^2(k_0)\left(\frac{k}{k_0}\right)^{n_s-1},
\end{equation}
where $ \Delta_{\cal R}^2(k) $ is the power spectra   of $ \cal{R}
$, the curvature perturbation in the comoving gauge, and $n_s$ is
the slope of the scalar power spectrum. The tensor perturbations
are given by:
\begin{equation}
\Delta_T^2(k)=\Delta_T^2(k_0)\left(\frac{k}{k_0}\right)^{n_t},
\end{equation}
where $ \Delta_T^2(k) $ is the power spectra of gravitational
waves from inflation and $n_t$ is the slope of the gravitational
wave power spectrum. The amplitude terms are evaluated at the
pivot wave number, $ k_0=0.05 \rm{Mpc}^{-1} $. To parameterize the initial
perturbations we use three inflationary parameters: $ A $, a
constant of order unity which is proportional to the amplitude of
the initial scalar perturbations, $ n_s $, the tilt of the power
spectra of the initial scalar perturbations and $ r $, the ratio
of tensor to scalar perturbations.  These are the parameters used
in the analysis of the {\it WMAP} data \cite{WMAP:pvalue}. We do not consider
here the running of the spectral index, $n_s'$.
%In previous versions of CMBFAST, the scalar perturbations have
%been parameterized by $ \psi $, the Newtonian scalar metric
%perturbation \cite{SZ}.  The normalization of the $ \psi $ power
%spectrum was then chosen so that the radiation power spectra, $
%C_{\ell} $, was the correct amplitude in units of $ \mu K^2 $.
%$ A $ is then the factor that this output needs to be scaled by to
%fit the observational data.
The exact relationship between $ A $ and  $ \Delta_{\cal R}^2(k_0)
$ is derived in Verde et al (2003; equation (32)). The tensor-to-scalar
ratio is defined as:
\begin{equation}
\label{eq_r} r=\frac{\Delta_T^2(k_0)}{\Delta_{\cal R}^2(k_0)}.
\end{equation}
Note that a number of different definitions are used in the
literature.  The most common alternatives are to define $ r $ in
terms of the Newtonian potential:
\begin{equation}
\label{eq_r2} r_{\psi}=\frac{\Delta_T^2(k_0)}{\Delta_{\psi}^2(k_0)},
\end{equation}
so that $ r_{\psi} \simeq (5/3)^2 r $, or in terms of the CMB radiation
quadrupoles:
\begin{equation}
r_Q=\frac{C_2^{T}}{C_2^{S}}.
\end{equation}
The relation between $ r $ and $ r_Q $ depends on the cosmological
parameters used in the model \cite{TW}.

The cosmological parameters we shall consider are $ \{
\Omega_bh^2, \Omega_mh^2, h, \tau \} $, where $h$ is the Hubble
constant in units of $100 \, \rm{km}s^{-1}\rm{Mpc}^{-1} $, $
\Omega_b $ is the energy density of baryons, $ \Omega_m $ is the
total matter density and $ \tau $ is the optical depth to the last
scattering surface. Again, these are the parameters chosen for the
{\it WMAP} data analysis \cite{WMAP:pmethod}. The full set of parameters, and
is then:
\begin{eqnarray*}
\lefteqn{\{ \Omega_bh^2, \Omega_mh^2, h, \tau, n_s, r, A  \}  =}  \\
  & & \{ 0.0224, 0.135, 0.71, 0.17, 0.93, 0.01, 0.83  \}. \nonumber
\end{eqnarray*}
The values of these parameters are taken from the best-fit {\it WMAP} model \cite{WMAP:pvalue}.
Although we do not include $n_s'$ in our analysis we note that these
best-fit parameters from a model which inculdes a non-zero $n_s'$.
For this parameter set the relation between the r and $ r_Q $ is
approximately $ r_Q \approx 2.8r $ and $ \Delta_{\cal
R}^2(k_0)=2.45 \times 10^{-9} $. The values of these parameters are
taken from the best fit {\it WMAP} model \cite{WMAP:pvalue} except for $ r $,
which cannot be well constrained by this data set.  The current
upper limit on $ r $ is about $ 0.36 $ \cite{LL}\footnote{Note that our value 
differs from the value given in this reference as we use a different 
value for the pivot wavenumber, $ k_0 $}. The lowest
possible $ r $ which can be detected is of the order of $ 10^{-4}
$ \cite{KS} due to noise left over from the removal  of the
gravitational lensing signal from the B-mode spectrum.  To reflect
this range of possible values we perform the calculations, which
have strong dependence on $ r $, at two different values, $ r=0.01
$ and $ r=0.1 $.

\section{Survey design} \label{sec:areaopt}
\subsection{Method}
The optimization of the survey area for a ground based measurement
of the CMB polarization has been addressed previously \cite{JKW}
in the context of  making a detection. We extend this work by
considering the criteria for a measurement of the
polarization spectra including the
effects of gravitational lensing and foregrounds.

The main aim of a polarization experiment is to make measurements
of the three polarization power spectra, $ C_{\ell}^{TE} $, $
C_{\ell}^{EE} $ and $ C_{\ell}^{BB} $, with the highest possible
precision.  The error in the
measurement of the power spectra is determined by two conflicting
factors. For a fixed total observing time, the integration time
per unit area (or pixel) is inversely proportional to the total
area; a smaller map will therefore result in a lower pixel noise.
However, for a smaller map there are fewer independent modes from
which to measure each multipole (i.e. the averaging in equation
(\ref{eq:multipole}) will be made over fewer values of $ m $) and
so the sample variance will increase.

To quantify these effects, we choose a single parameter for which
to evaluate the Fisher matrix, $ A^{\!X} $, the amplitude of each
power spectrum. For a single parameter the variance in the
measurement of this parameter, $ (\Delta A^{\!X})^2 $, is then
given by $ 1/F_{A^X \!A^X} $. From equation (\ref{eq:fishs}) the
error in $ A^{\!X} $ is:
\begin{equation}
(\Delta A^{\! X})^2=\left(\sum\limits_{\ell} \frac{1}{
 (\Delta C_{\ell}^X)^2}\frac{(C_{\ell}^X)^2}{(A^{\! X})^2}\right)^{-1},
\end{equation}
where $ (\Delta C_{\ell}^X)^2 $ for each power spectrum are  given
by the diagonal elements of the power spectrum covariance matrix
in equation (\ref{eq:PScov}).  We then define a figure of merit
parameter as the signal to noise ratio in the measurement of each
power spectrum, SNR, which is given by:
\begin{equation}
\label{eq:SNR} \rm{SNR}=\left(\frac{A^{\! X}}{\Delta
A^{\!X}}\right)= \sqrt{\sum_{\ell}\left(\frac{C_{\ell}^X}{\Delta
C_{\ell}^X}\right)^2}.
\end{equation}
To find the optimal area for a measurement of each power spectrum with a specfic experiment
we therefore need to find the area which gives the highest SNR
given a set of survey and instrument parameters.

This optimization procedure could be done for any cosmological parameter, or
combination of parameters. However for simplicity, and as a
prerequisite to the measurement of the polarization power spectra,
we will maximize the SNR for the amplitude. In principle, other
parameters for the survey or telescope could be left free, such as
the pixels size or beam width. In practice we find that the
smallest pixel/beam size is preferred, and so we set this to the
limit of a given experiment.

\subsection{\QU{} instrument parameters} \label{sec:QU}
As a specific example of a ground-based experiment we use the \QU{}
experiment. This enables us to fix the instrument parameters needed to determine the
pixel noise (equation \ref{eq:pnoise}) and the allowed multipole range. These parameters
are given in Table \ref{tb:inst}. A detailed description of \QU{} is given in Church et al. (2003).

The maximum multipole which can be covered is limited by the beamsize as discussed in Section 
\ref{sec:formal}. If no other effect needs to be taken into consideration the minimum multipole, 
$\ell_{min}$, would be determined from the survey area, $\ell_{min}=\pi/\Theta $. However,
for a ground-based experiment the lower-$\ell$ cut-off is also limited by the stability of the atmosphere.
This will limit the maximum scan which can be used and hence the largest angle on the 
sky over which a correlation can be made. For a perfect polarization experiment, this would not
be an issue, as the unpolarized atmospheric fluctuaions would not be detected in the polarized data.
However, instrumental effects will cause a fraction of the unpolarized (common mode) signal
to be to be present in the polarized signal. The atmosphere at the South Pole is exceptionally
stable \cite{atm} and the \QU{} instrument has been designed is such a way that these effects will
be minimized, so we estimate that a minimum $\ell$ of $25$ can be reached. The minimum $\ell$ used in 
equation (\ref{eq:SNR}) will then be $\ell_{min}= {\rm max}(\pi/\Theta,25)$. Although it is
possible for \QU{} to make total power measurements, there is no mechanism for removing the
atmospheric noise from the resulting data and so we assume that \QU{} would not be able to
produce temperature maps.

We estimate the total observing time
by assuming that \QU{} will observe at the South Pole for two years during the
austral winter (six months per year) for 22 hours each day and assuming
that $ 20 $ per cent of this total time will be lost due to bad
weather, instrument maintenance and calibration time. These
estimates are based on the experiences of the DASI team at the
South Pole site \cite{DASI}. This gives a total time spent
observing on the CMB of $ 3210 $ hours per year. The maximum useable
patch of sky is about $ 1000 $
deg$^2$, limited by available sky visible from the survey site,
and major foreground contamination from the Galactic plane (see
Fig. \ref{fig:Kenfg}). We therefore restrict the analysis to
areas below this maximum survey size.

It is important to note that to measure both the Q and U
Stokes parameters, each pixel must be measured with the detector in at least
two different orientations with respect to the sky.
For \QU{} this will be achieved by rotating
a half-wave plate so that both Q and U can be measured by each detector.
This halves the total integration time available for each Stokes parameter
when making a polarized measurement.
\begin{table}
\caption{\label{tb:inst} Expected \QU{} instrument parameters}
\begin{tabular}{|l|c|c|}\hline
 Frequency (GHz) & 100 & 150 \\
 Number of bolometers & 24 & 38 \\
 Angular resolution (arcmin) & 6.3 & 4.2 \\
% NET per bolometer &  270  & 300 \\ \hline
 NET per bolomete$\mbox{r}^{a} $ ($ \mu \mbox{K} s^{1/2} $) &  270  & 300 \\ \hline
\end{tabular}

\medskip

a - The definition of sensitvity for a polarization sensitive bolometer is discussed in
Appendix A. \\ 
\end{table}

\subsection{Results}

We have applied the above procedure to a model \QU{} experiment.
We consider three different cases:
\begin{enumerate}
\item a measurement of the EE spectrum,
\item a measurement of the BB spectrum including the lensing
    component as part of the signal we wish to measure,
\item a measurement of just the BB GW spectrum including the
    lensing signal as an extra source of noise.
\end{enumerate}

The results for the \QU{} parameters
are presented in Table \ref{tb:area}, for TE, EE and BB spectra, for the case
of foregrounds, and without foregrounds. The latter is of interest
if the foregrounds are well enough understood to be subtracted
from the signal or if a patch of sky with very low foreground variance can
be found, as discussed in Section \ref{sec:fg} 

\begin{figure}
\psfig{file=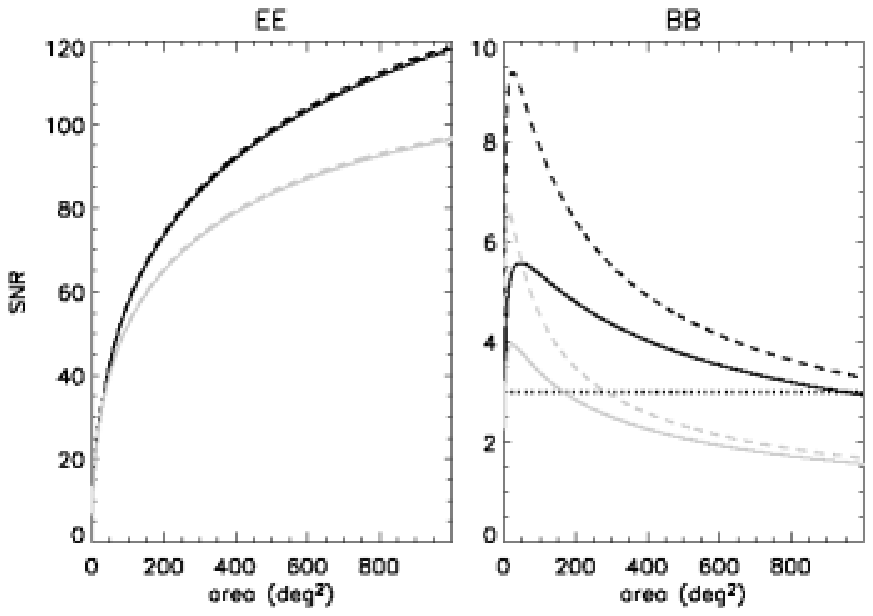,width=9.0cm,height=8.0cm}
\caption{Variation of SNR with survey area for the EE (left) and
total BB (right) power spectra with foregrounds (solid  lines) and
without foregrounds (dashed lines) for an observing time of 1 year
(light) and 2 years (dark) for r=0.01} \label{fig:areaEE}
\end{figure}

Fig. \ref{fig:areaEE} shows how the SNR varies with area for EE
and BB spectra. From Fig. \ref{fig:areaEE} (left) it can be seen
that a ground-based polarization experiment can make a good
measurement of the EE spectrum, even if the foreground
contamination is not well understood. The SNR is close to 100 for
survey areas over $300$ deg$^2$. Below this size the
SNR falls rapidly to zero. For an experiment with the sensitivity and multipole coverage
of \QU{}, an E-mode survey is sample-variance-limited, so that a
larger area is preferable ($> 1000 $ $\rm{deg}^2 $) for
statistical purposes. Given that an E-survey will have high
signal-to-noise per pixel, a high-resolution polarization map of
the surveyed area is also possible (see Section \ref{sec:maps}), allowing for removal of point
sources as discussed in Section \ref{sec:fg}.

Fig. \ref{fig:areaEE} (right) shows the SNR for a two-year B-mode
survey. The dotted line is for a SNR=3 which is the minimum SNR that
can be considered as an actual detection of the signal. Unlike the E-mode survey,
the B-mode survey is detector noise limited as the signal is much lower.
With foregrounds the SNR sharply peaks at ${\rm SNR} \approx 5.5$ for much smaller areas, 
around $50$ deg$^2$, where the lensing
signal dominates. As the survey area increases the noise per pixel
increases and the overall SNR drops. If we can remove foreground contamination
then the maximum SNR increases to a value of 9 and the optimal area is
slightly reduced, as shown in Table \ref{tb:area}.

This different behaviour between the E and B-mode surveys with
increasing area makes the simultaneous optimization of both
measurements difficult. One compromise is to use the break in SNR
of the E-mode survey at around 300 $ \rm{deg}^2 $. Although this
is sub-optimal for both surveys, the drop to ${\rm SNR}= 90$  for
the E-survey is minimal and ${\rm SNR}=4.5$ for the B-mode survey is
still a strong detection. An alternative would be to split the
survey in two, one large, one small, halving the integration time
for each survey. As shown in Fig. \ref{fig:areaEE} for a single
year of integration it is still possible to detect the B-mode signal
if we concentrate on a small area of sky ($\sim 20$
$\rm{deg}^2$).  For a single year the EE SNR also does not drop
significantly if an area larger than about 500 $ \rm{deg}^2 $ is
chosen.

\begin{figure}
\psfig{file=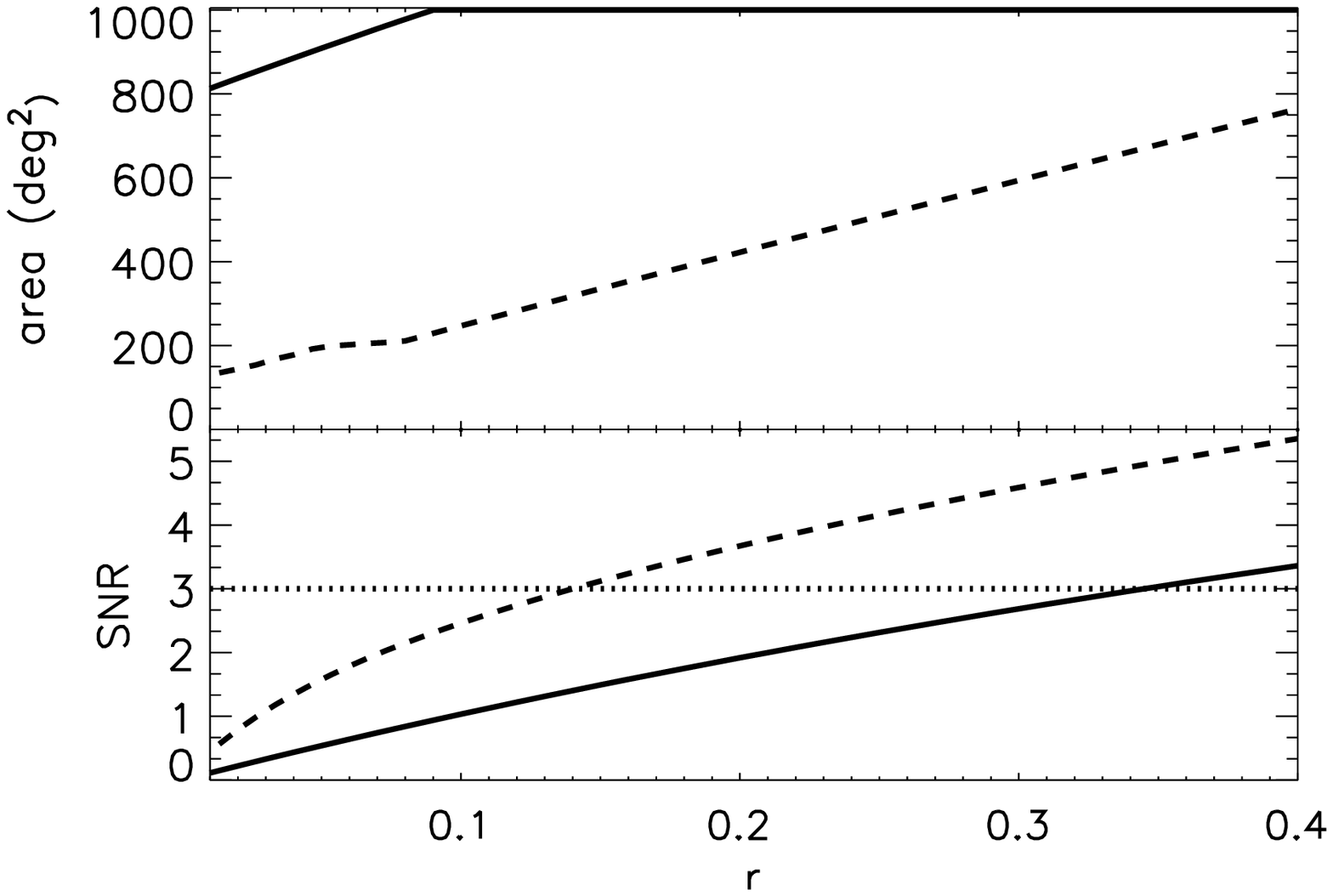,width=7.5cm,height=7.0cm}
\caption{Variation of the survey optimal area (upper plot) and achievable SNR (lower plot)
for a measurement of
the GW signal with \QU{} as a function of $ r $ with (solid line) and without (dashed line) foregrounds.}
\label{fig:areaGW}
\end{figure}

From Table \ref{tb:area} it is evident that \QU{} cannot detect the GW B-mode
component unless the tensor to scalar ratio is larger than the values considered so far.
We have therefore extended the calculation to higher values of $r$ up to the current 
upper limit. Fig. \ref{fig:areaGW} shows how the optimal area for a measurement of the GW signal with \QU{} 
varies with $ r $. The optimal area changes significantly as $r$ increases.
For the foreground model assumed here it is only possible to detect the GW
signal for $ r $ greater than 0.35.  
However, for the large areas which are best for detecting this high GW signal, the SNR
for the total B-mode signal drops significantly. It is therefore not possible
to pursue both science goals simultaneously. However, if the foreground comtamination can be 
completely removed, the lowest detectable value of $ r $ drops to 0.14. The optimal area
also decreases as the detector noise becomes the dominant factor. In the no foreground
case it would be possible to detect the GW signal using the 
$ 300$ $\rm{deg}^2$ survey discussed above.    
If the GL signal can be removed, the GW signal becomes slightly easier to detect, but only
if the foregrounds can be subtracted as
the combined dust and synchroton contamination (Fig. \ref{fig:fgpower}) is larger 
than the GL signal over most of the multipole range which can be covered
from the ground.  

The effects of the mixing of E and B modes due to partial sky coverage will
not significantly influence the results found here. Bunn (2002) finds that the
mixing will only have a large effect for the B-mode signal on the scale of 
the survey size. If we use a $300$ $\rm{deg}^2$ patch the GL B-mode signal will therefore
not be affected. For a detection of the GW signal this effect will become more important. 
However, Lewis, Challinor \& Turok (2002) discuss this problem and calculate the minimum detectable $r$
as a function of survey size. They find that for the large surveys (greater than $50^{\circ}$) the minimum
value is not changed if the mixing effects are included. For the areas discussed here
the GW results will therefore not be influenced by E-B mixing if an optimal method is used to separate
the E and B modes.    

\begin{table}
\caption{\label{tb:area} SNRs for the optimal survey areas for each of the power spectra for a two-year integration time with \QU{}}
\begin{tabular}{|c|c|c|c|c|c|c|c|}
\hline  \multicolumn{2}{c|}{Spectrum} & TE & EE & \multicolumn{2}{c|}{BB} & \multicolumn{2}{c|}{$GW$}   \\
\multicolumn{2}{c|}{$r$} & 0.01 & 0.01 &  0.01 &  0.1   &  0.01  & 0.1    \\
\hline  \multicolumn{2}{c|}{area$^a$ /$ \rm{deg}^2 $} &  1000+ & 1000+ & 46 & 50 & 813 & 1000+ \\
\multicolumn{2}{c|}{area$^b$ /$ \rm{deg}^2 $} &  1000+ & 1000+ &  24 & 26 & 126 & 247 \\
\hline  \multicolumn{2}{c|}{SNR$^a$} & 31 & 118 & 5.6 & 5.7 & 0.1 & 1.0 \\
\multicolumn{2}{c|}{SNR$^b$} & 31 & 119 & 9 & 9 & 0.4 & 2.5 \\
\hline
\end{tabular}

\medskip
a - including astrophysical foregrounds \\
b - without astrophysical foregrounds

%\medskip
%one year integration time \\
%\begin{tabular}{|c|c|c|c|c|c|c|c|c|c|c|c|}
%\hline  \multicolumn{2}{c|}{Power spectra} & TE & EE & \multicolumn{2}{c|}{BB} & \multicolumn{3}{c|}{$GW^a$} & \multicolumn{3}{c|}{$GW^b$}   \\
%         &                               &    &    &   r=0.01   &  r=0.1       &  r=0.01        &   r=0.1   &  r-0.3  &     r=0.01    &     r=0.1  & r=0.3    \\
%\hline  area & with foregrounds &  1000+ & 1000+ & 16 & 18 & 134 & 212 & 459 & 191 & 270 & 505 \\
%  & no foregrounds &  1000+ & 1000+ &  12 & 13 & 44 &  126 & 252 & 85 & 165 & 297 \\
%\hline  SNR &  with foregrounds & 30 & 97 & 4.7 & 4.8 & 0.1 & 0.1 & 2.3 &  0.1 & 0.1 & 2.3 \\
%  &  no foregrounds & 31 & 97 & 7 & 7 & 0.4 & 1.9 & 3.4 &  0.3 & 1.7 & 3.2 \\
%\hline
%\end{tabular}

%\medskip
%a - assuming that all of the lensing B-mode signal can be perfectly removed from GW component\\
%b - including all of lensing signal as an extra source of noise
\end{table}
To estimate the SNR for the TE spectrum we assume that a \QU{}
map could be combined with the portion of the expected four-year
{\it WMAP} data covering the same area of sky. The results are
shown in Table \ref{tb:area}. As with the EE spectrum, the measurement is
sample-variance-limited and the largest possible area of $1000$
deg$^2$ is best. The SNR also drops sharply if the survey area
becomes too small ($\le$ $100$ deg$^2$). However, the \QU{} TE
measurement is limited by the resolution and sensitivity of the
{\it WMAP} map and suffers more heavily from sample variance than
the smaller EE signal. The SNR with which this spectrum could be
measured by \QU{} is therefore smaller than the EE SNR. The TE
spectrum has also already been measured in this multipole range by
{\it WMAP}. It is therefore more useful to optimize a ground-based
survey for a measurement of the EE and BB spectra.

We have also investigated the effect of increasing the minimum
$ \ell $ value used in the calculation.  For the TE, EE and total 
BB spectra, an increase in the minimum $ \ell $ from 25 to 100 has a
negligible effect, as most of the power in these spectra is from the 
higher multipoles. However, as would be expected, increasing the 
minimum $ \ell $ does affect the GW B-mode detection. If the minimum $ \ell $ 
is increased to 100 the GW is no longer detectable below
the current upper limit of $ 0.36 $.

\section{Deep maps of the CMB polarization}
\label{sec:maps}
\begin{figure*}
\epsfxsize=6.5cm
\epsfysize=6.5cm
\subfigure{\epsffile{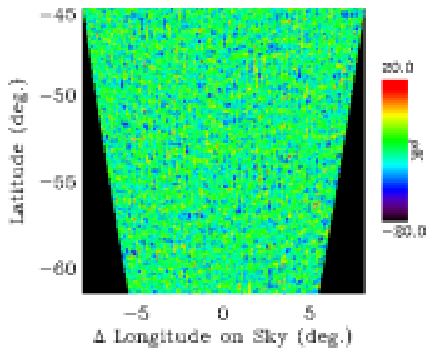}}
\epsfxsize=6.5cm
\epsfysize=6.5cm
\subfigure{\epsffile{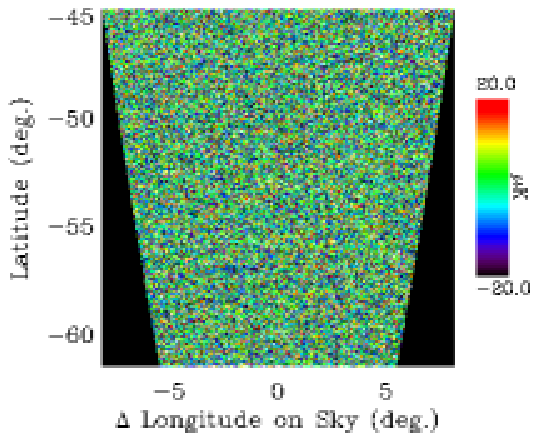}}
\caption{Simulated polarization maps from \QU{} (left) and Planck
(right), showing a $ 300$  $\rm{deg}^2 $ field of Q-mode anisotropies at 150
GHz. All of the structure in the high signal-to-noise \QU{} map is
real signal, while the Planck map (shown with the same pixelization) has
much lower signal-to-noise per pixel. Foregrounds and other
systematic effects are not included in either map.} \label{fig:map}
\end{figure*}

For a ground-based experiment it is not possible to make
observations of the whole sky due to the limited sky coverage 
available from the ground. Although this is a disadvantage in terms of
multipole coverage at low $ \ell $, by making a deep integration of a small
region of sky it is possible to make maps with a very high signal-to-noise ratio.
This allows more precise measurements to be made on small angular scales.
It will also improve the ability of the experiment to remove low-lying
systematic effects which would not be detectable in observations of lower
signal-to-noise. We illustrate the difference between \QU{} and the {\it Planck}\footnote{http://www.astro.esa.int/SA-general/Projects/Planck/}
satelllite mission in Fig. \ref{fig:map} using simple simulations of the Q Stokes
parameter with noise appropriate to each experiment. While {\it Planck} will
cover much more sky than \QU{}, the \QU{} observations would be at higher signal-to-noise
than the average of those made by {\it Planck}\footnote{It is noted that the average noise over the
entire sky was used for {\it Planck}. While {\it Planck} will cover some regions, namely the 
Ecliptic poles, more deeply, these regions are in general not the best in terms of foregrounds,
and the mean noise away from these regions will be correspondingly worse}. 
This will allow \QU{} to limit systematic effects in the experiment to an unprecented level. This makes
the \QU{} approach highly complementary to that of {\it Planck}, which 
would have lower average sensitivity, but good statistics over the entire sky. The deep maps will
also provide new information on the techology used by \QU{} and on polarized foregrounds, which will be
crucial for the design of future CMB experiments.
%This would allow \QU{}
%to make a better noise characterization, although {\it Planck} will have correspondingly better
%statistics as it will observe many regions of this size. Many of the technologies which would be 
%piloted by \QU{} will be used in the next generation of instruments, and so this information
%will help to guide the design of future more sensitive experiments.

Using a smaller region of sky is also an advantage in terms of foregrounds, as it is possible
to target the most useful patches of sky, without spending valuable integration time on regions
which will ultimately be left unused in cosmological analyses.

Finally, it is possible tailor the size of the region observed to optimize for
a particular science goal. As was described in Section \ref{sec:areaopt}, this is 
especially important for searches for the faint B-mode signal.

\section{Power spectra estimation} \label{sec:spec}

\begin{figure*}
%\centering
\hspace{0.7cm}
\psfig{file=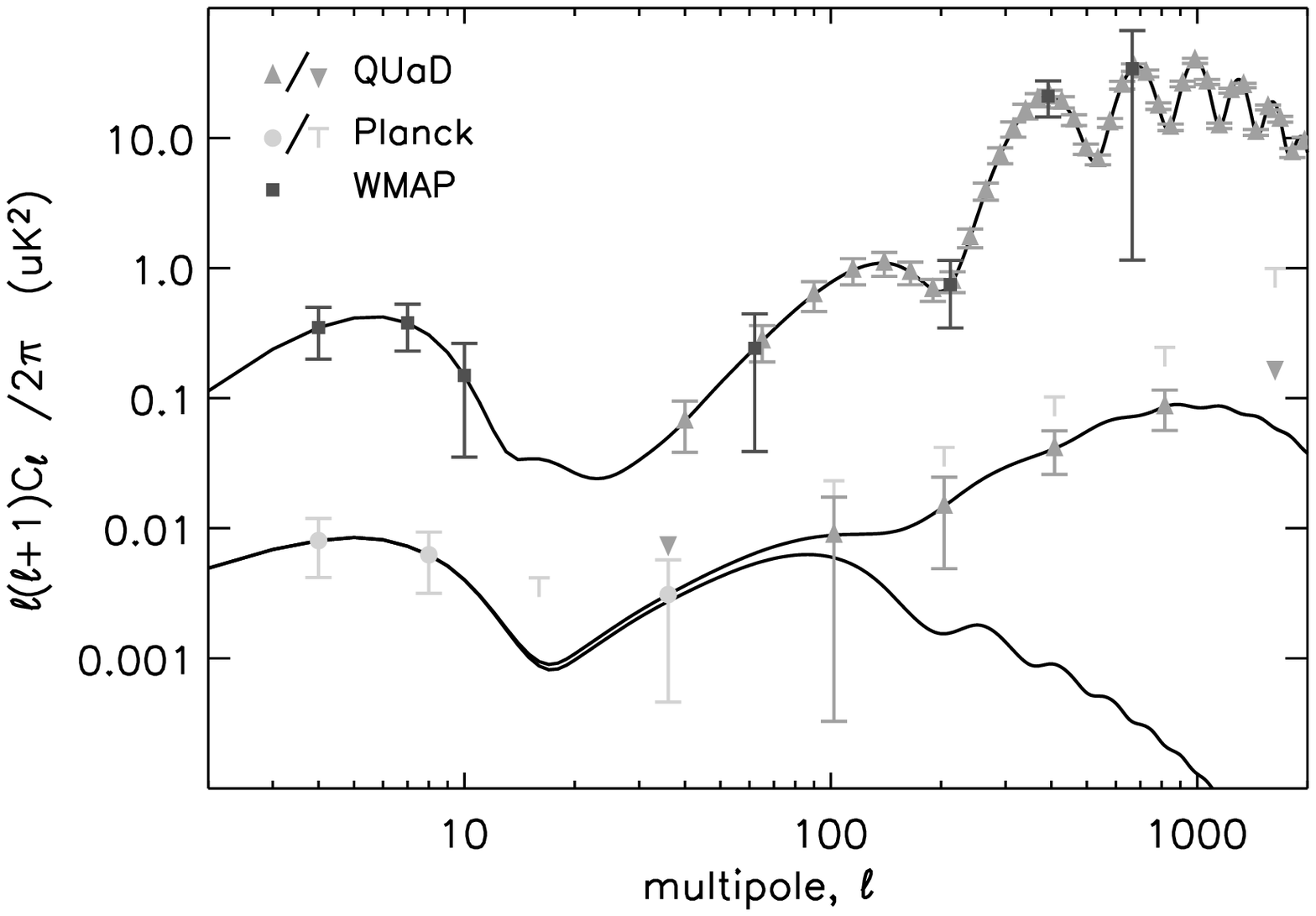,width=12.0cm,height=8.5cm}
\caption{Predicted measurements of the polarization
power spectra achievable with the current generation of satellite
(four-year {\it WMAP} and two-year {\it Planck}) and ground-based (two-year \QU{})
experiments for $ r=0.1 $. The error bars show detections above
the one-sigma level and the free symbols show upper limits. The errors
include a contribution from astrophysical foregrounds. For
clarity we do not show the {\it Planck} measurements of the EE
power spectrum. These will be of a similar sensitivity to the
\QU{} measurements, but will cover different $ \ell $ ranges, as
indicated by the points on the B-mode spectrum.} \label{fig:ps}
\end{figure*}

\begin{table}
 \caption{\label{tb:sat} {\it Planck} and {\it WMAP} instrument
parameters. For {\it WMAP} we use only the highest two frequency
channels as the other channels are used mainly to constrain
foreground contributions.}
\begin{tabular}{|c|c|c|c|c|c|c|}
\hline    & \multicolumn{4}{c|}{\it Planck} & \multicolumn{2}{c|}{{\it WMAP}} \\
\hline  frequency /GHz & 40 & 70 & 150 & 220 &  70 & 90  \\
  NET /$\mu Ks^{1/2}$ & 220 & 300 & 80 & 120 &  1521 & 2071 \\
  Beam size /arcmin & 24 & 14 & 7 & 5 & 20  &  13\\
  Detector number & 6 & 12 & 8 & 8 & 8 & 16 \\
\hline
\end{tabular}
\end{table}

By statistically averaging over the polarization signal, the
polarization power spectra may be estimated. Fig. \ref{fig:ps}
compares the expected band-averaged power spectra results and
multipole coverage of a 300 deg$^2$, two-year ground-based
experiment, \QU{}, an all-sky four-year satellite, {\it WMAP}, and
the {\it Planck } satellite mission. These predictions are based
on equation (\ref{eq:XYXY}), with the parameter the power in a
pass-band of width $\Delta \ell$. We include all of the power
covariances and effects of foreground emission outlined in Section
\ref{sec:fg}. The instrument parameters used in each experiment are
given in Tables \ref{tb:inst} and \ref{tb:sat}.

From this analysis we find that \QU{}
can make a high-significance measurement of the EE-power spectrum,
as suggested by the high SNR ($\sim 90$) found during
optimization in Section \ref{sec:areaopt}, over a multipole range
from $\ell=25$ to $\ell=2500$. The polarization acoustic
oscillations are well sampled, with a resolution of $\Delta \ell
\approx 20$. The lower modes are not sampled due to the limited
survey area. In particular the re-ionization peak at $\ell=7$ is
only detected by a satellite mission.

In addition there is a good detection of the BB-power spectrum
from $\ell=25$ to $\ell=1000$. Power is binned logarithmically to
increase the signal-to-noise per bin. The most significant bin is
at $\ell=1000$, at the peak of the gravitational lensing (GL)
contribution to the BB-spectrum. If the foreground
contamination can be significantly reduced, a direct detection
of the GW contribution to the B-mode power spectrum could be made
at around $ \ell =100 $. Again
the low-$\ell$ modes are not accessible to a ground-based survey,
but can be complementarily detected by an all-sky satellite
mission.

With both temperature and polarization data available the TE-cross
power spectra may also be estimated so that a cross-check can be
made with other measurements of this signal.

With such high-resolution polarization information available it is
interesting to see what effect a ground-based survey will have on
cosmological parameters.

\section{Parameter Estimation}
\label{sec:param}
\input{param.tex}

\section{Summary}
\label{sec:discuss}

In this paper we have investigated the science goals achievable
with the forthcoming generation of ground-based CMB polarization
experiments. We have set out a Fisher information matrix formalism
that takes into account the combination of different temperature
and polarization surveys, and includes foreground contamination.
We have argued that ground-based polarization experiments can
reach the high sensitivities required by making a deep integration
on a small patch of sky. By preferentially selecting regions of sky
with low foreground variance it will also be possible for 
a ground-based experiment to reduce further  the foreground contamination.

Taking the proposed \QU{} South Pole experiment as our model survey, we
have optimized the survey area and shown that a 300 $ \rm{deg}^2 $
survey is a good compromise between a sample-limited E-mode survey
and a detector-noise limited B-mode survey. Below 300 $ \rm{deg}^2 $ the
SNR for the E-mode survey drops rapidly, while above this a
detection of the (gravitational lensing component of the) BB-power
spectrum becomes unfeasable. With such high signal-to-noise per
pixel in the E-mode survey, deep imaging maps of the CMB
polarization field can be made. Statistically averaging the data
allows a high-significance measurement of the EE-power spectrum
over a range of multipoles from $\ell=25$ to $\ell=2500$, with
good sampling of the acoustic oscillations. The gravitational
lensed component of the BB-power spectrum can also be detected
with good signal-to-noise. If it is possible to reduce the foreground
contamination the gravitational wave could also be detected for
$ r \geq 0.14 $.

Combining a two-year \QU{} experiment with a four-year {\it WMAP} all-sky
survey allows a better measurement of cosmological parameters to be made compared
to that possible from {\it WMAP} data alone.
Most parameters can be improved by a factor two. If the foreground contamination can be reduced the tensor-to-scalar
ratio will be dramatically improved by up to a factor of six.
With such improvements, strong constraints can be placed on the
potential of the inflaton field. Only the degeneracy between the
amplitude of fluctuations, $A$, and the optical depth to
re-ionization, $\tau$, are not significantly improved, as this
requires large scales only accessible to a satellite.

In conclusion we find that if the necessary sensitivity and control
of systematics can be achieved,
a ground-based CMB polarization
experiment such as \QU{} can make a major contribution to the study of CMB polarization
power spectra and cosmological parameters.

\section*{Acknowledgements}

MB would like to acknowledge a departmental grant from the
University of Wales, Cardiff. ANT thanks the PPARC for an Advanced
Research Fellowship. The US contribution to this work is supported by the 
National Science Foundation under grants 9987360 and 0096778.

\appendix
\section{Sensitivity definitions for CMB polarization experiments}
\input{appendix.tex}

\end{document}

%% file: author_list.tex
\author[M. Bowden et al]
{M. Bowden,$^1$\thanks{Melanie.Bowden@astro.cf.ac.uk} A. N. Taylor,$^2$\thanks{ant@roe.ac.uk}
K. M. Ganga,$^3$\thanks{kmg@ipac.caltech.edu} P. A. R.
Ade,$^1$ 
\newauthor J. J. Bock,$^{5,6}$ G. Cahill,$^7$
J. E. Carlstrom,$^8$ S. E. Church,$^4$ W. K. Gear,$^1$
\newauthor J. R. Hinderks,$^4$ W. Hu,$^8$ B. G. Keating,$^6$
J. Kovac,$^{6,8}$A .E. Lange,$^6$ E. M. Leitch,$^8$
\newauthor B.
Maffei,$^1$ O. E. Mallie,$^1$ S. J. Melhuish,$^1$ J. A.
Murphy,$^7$ G. Pisano,$^1$ L. Piccirillo,$^1$ \newauthor
C. Pryke,$^8$ B. A. Rusholme,$^4$ C. O'Sullivan,$^7$
K. Thompson$^4$\\
$^1$Department of Physics and Astronomy, University of Wales, Cardiff, PO Box 913, Cardiff, CF24 3YB\\
$^2$Institute for Astronomy, University of Edinburgh, 
Royal Observatory, Blackford Hill, Edinburgh, EH9 3HJ\\
$^3$Infrared Processing and Analysis Center, California Institute of Technology, Pasadena, CA 91125\\
$^4$Department of Physics, Stanford University, Stanford, CA 94305\\
$^5$Jet Propulsion Labratory, 4800 Oak Grove Dr., Pasadena, CA 91109\\
$^6$Division of Physics, Math and Astronomy, California Institute of Technology, Pasadena, CA 91125\\
$^7$Experimental Physics Department, National University of Ireland, Maynooth, Co. Kildare, Ireland\\
$^8$Department of Astronomy and Astrophysics, Department of Physics, Enrico Fermi Lab, University of Chicago, \\
5640 South Ellis Avenue, Chicago, IL 60637}

%% file: foreground.tex
\subsection{Including foregrounds into the formalism}
\label{sec:fg1}
The signal measured from the sky will contain not only a component
from the CMB, but also a contribution from
astrophysical foregrounds. The CMB signal is independent of the
wavelength of the observation, but the signal from most
foregrounds is expected to be frequency dependent.  By observing
in a number of different frequency channels it is therefore possible
to reduce the total foreground contamination by optimally combining the
signal from different frequency channels. It may also be possible to 
use the multiple frequency information to remove some of the
foreground contamination from the signal 
(e.g. Maino et al. 2002, Hobson et al. 1998).

The effect of
observing over multiple channels needs to be taken into account in
the Fisher matrix formalism described in the previous Section.  If
we ignore foregrounds and consider only detector noise, we can
simply replace the noise terms in equation (\ref{eq:XYXY}) by an
inverse variance weighting of the noise in each channel, $
N_{\ell,c} $:
\begin{equation}
\label{eq:weight} N_{\ell} = \left( \sum_c
\frac{1}{N_{\ell,c}}\right)^{-1}.
\end{equation} 
By choosing this weighting scheme at each multipole we combine the
signals by giving the most weight to the channels with the
smallest detector noise.

We include the effect of foregrounds by treating the foregrounds
as an extra source of noise with power spectra $ N_{\ell}^{fg} $
for each different power spectra in each frequency channel. This gives us
the maximum possible foreground contamination i.e. the contamination assuming
that no foreground removal will be attempted.
However, unlike the detector noise, the foregrounds will be
correlated between power spectra and between frequency channels.
To include these correlations we follow the technique developed in
Tegmark et al. (2000, hereafter T00).  We define a $ 3F \times 3F $ noise matrix,
$ {\cal{N}}_{\ell} $, for each multipole, where $F$ is the number of frequency
channels in the experiment:
\begin{equation}
\label{eq_N}
{\cal{N}}_{\ell}=
\left(\begin{array}{ccc}
{\mathbf N}_{\ell}^{TT} & {\mathbf N}_{\ell}^{TE} & 0 \\
{\mathbf N}_{\ell}^{TE} & {\mathbf N}_{\ell}^{EE} & 0 \\
    0         &      0        & {\mathbf N}_{\ell}^{BB}
\end{array}\right),
\end{equation}
where each component of this matrix, $ {\mathbf N}_{\ell}^{XX'} $,
is an  $ F \times F $ matrix giving the variances and covariances
of the noise in the $F$ channels.  Each element in $
{\cal{N}}_{\ell} $ is the sum of the contribution from each
 of the possible foregrounds, $ {\mathbf N}_{\ell(k)}^{XX'} $
and the detector noise, $ {\mathbf N}_{\ell(det)}^{XX'} $:
\begin{equation}
{\mathbf N}_{\ell}^{XX'} = {\mathbf N}_{\ell(det)}^{XX'} + \sum_k
{\mathbf N}_{\ell(k)}^{XX'},
\end{equation}
where the sum over $ k $ is a sum over each of possible
foregrounds which could contribute to the signal. We define the $
3F \times 3 $ scan matrix, $\A$, where:
\begin{equation}
\label{eq:A}
\A= \left(\begin{array}{ccc}
{\mathbf e} & 0 & 0 \\
0 & {\mathbf e} & 0 \\
0 & 0 & {\mathbf e}
\end{array}\right),
\end{equation}
and $ {\mathbf e} $ is a column vector of height F  consisting
entirely of ones. If $ F=2 $ as would be the case for \QU{} (see Section \ref{sec:QU}) then:
\begin{equation}
\A= \left(\begin{array}{ccc}
1 & 0 & 0 \\
1 & 0 & 0 \\
0 & 1 & 0 \\
0 & 1 & 0 \\
0 & 0 & 1 \\
0 & 0 & 1 \\
\end{array}\right).
\end{equation}
The weighted noise for each polarization is  then obtained by
calculating the $ 3 \times 3 $ covariance matrix, $ \Sigma_{\ell}
$, where:
\begin{equation}
\Sigma_{\ell}=\left( \A^t  {\cal{N}}_{\ell} \A \right)^{-1} =
\left(\begin{array}{ccc}
N_{\ell}^{TT} & N_{\ell}^{TE} & 0 \\
N_{\ell}^{TE} & N_{\ell}^{EE} & 0 \\
    0         &      0        & N_{\ell}^{BB}
\end{array}\right).
\end{equation}
The terms, $ N_{\ell}^{XX} $ are  now the noise terms used in
equation (\ref{eq:XYXY}) to calculate the power spectra covariance
matrix.  If the noise is not correlated between T and E  and not
correlated between channels (as is the case if we include only
detector noise) then $ {\cal{N}}_{\ell} $ becomes diagonal and the
procedure is identical to the minimum variance weighting of
equation (\ref{eq:weight}).

In the last section we discussed how to combine a number of experiments by adding the 
Fisher matrices of independent patches of sky. For patches in which a number of 
experiments overlap, the required noise term, $ N_{\ell} $, can be calculated by 
considering a single experiment with channels at each of the different frequencies
used by this set of experiments. For this patch $ F $ will then become the  
total number of frequency channels in the combined survey. If any of 
the instruments used have channels which cannot measure either temperature or polarization,
then rows and columns corresponding to these channels should be
removed from the full noise matrix, $ {\cal{N}}_{\ell} $ and from the scan 
matrix, $\A$, in the relevant places. For example, \QU{} would not be able to
measure temperature information (see
Section \ref{sec:QU}). If we combine the two \QU{}
channels with another experiment measuring both temperature and
polarization, these two channels should be removed from the first row and first 
column of the matrix $ {\cal{N}}_{\ell} $ in equation (\ref{eq_N})
and the size of the vector $ {\mathbf e} $ in the first column of the matrix $\A$
in equation (\ref{eq:A}) should be reduced.

\subsection{Foreground models}

\begin{figure}
%\begin{minipage}{7cm}
\psfig{file=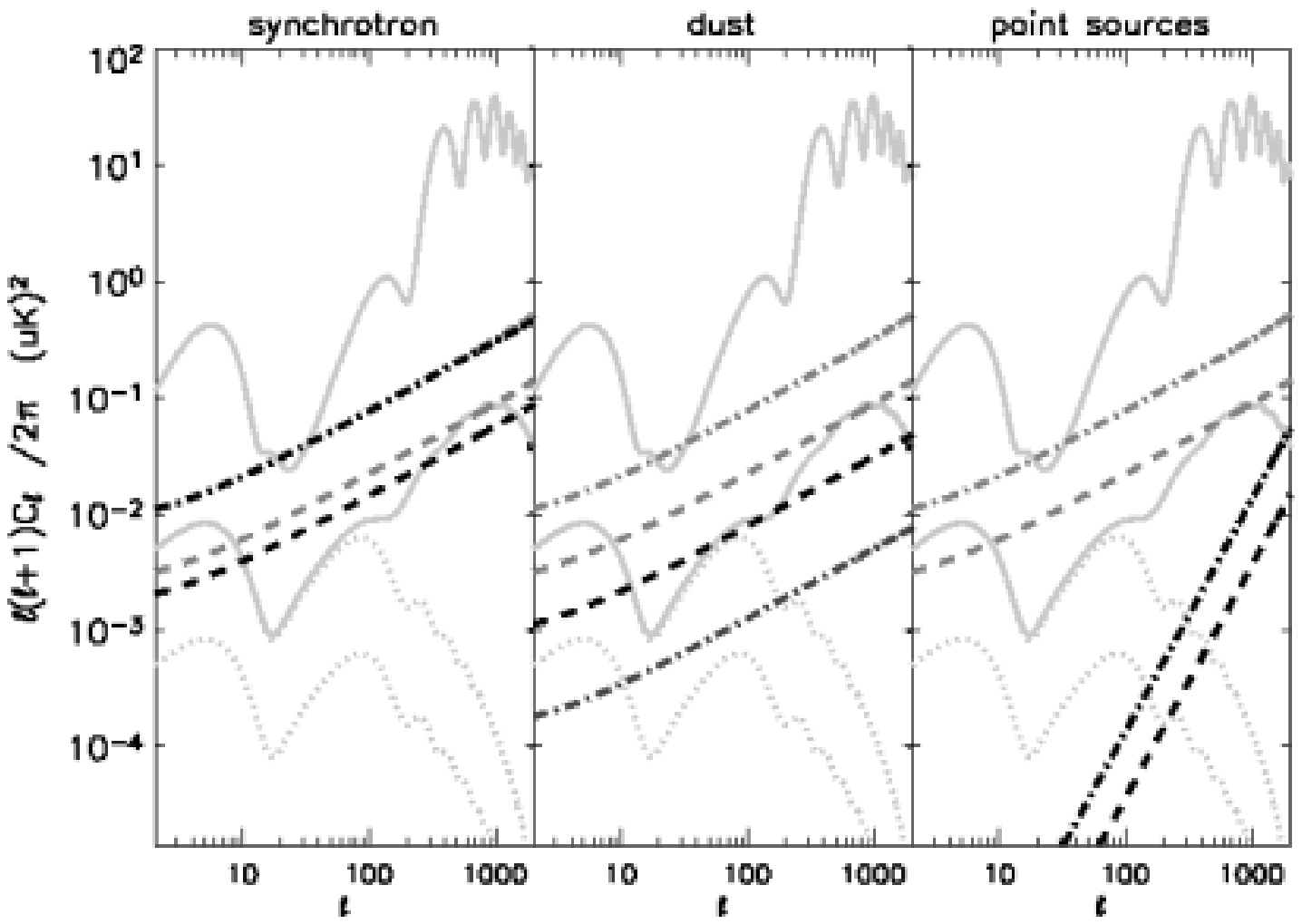,width=8.5cm,height=7.0cm}
%\end{minipage}
\caption{Models used for vibrating dust, synchroton emission and residual
point sources (black) compared to the EE and BB power spectra (light grey, solid).
The different lines show foreground models at 100 GHz (dash-dot) and 150 GHz (dash).
The total foreground power spectra are also shown on each plot (dark grey).
The GW component of the BB-spectra is shown for $r=0.1$ (upper dotted line)
and $r=0.01$ (lower dotted line). The other foregrounds are either unpolarized 
or can be neglected}
\label{fig:fgpower}
\end{figure}

\begin{figure}
%\begin{minipage}{7cm}
\psfig{file=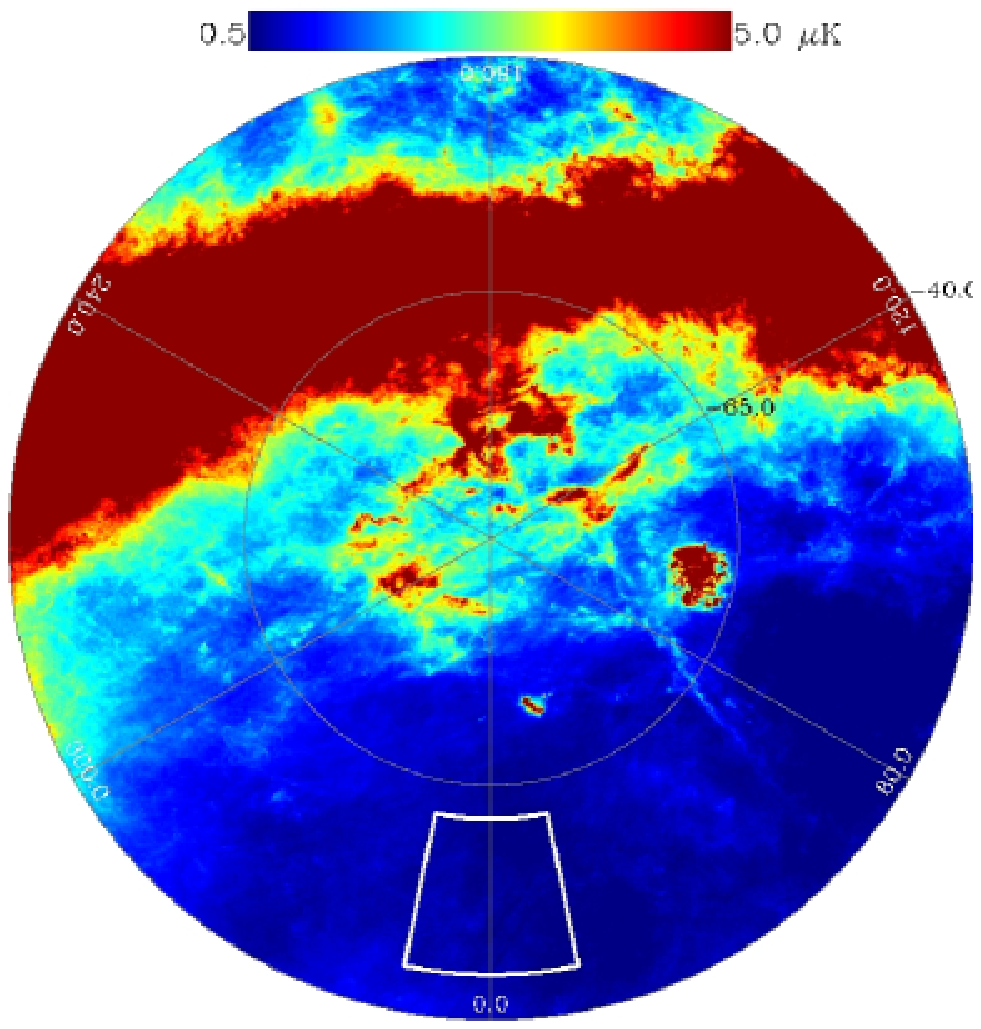,width=7.5cm,height=8cm}
%\end{minipage}
\caption{Equal-area zenithal projection showing foreground levels (dust
and synchrotron) at 150 GHz in regions which would be accessible to \QU. The
Southern Celestial Pole in located in the centre of the plot, and
declination -45 around the perimeter. To the bottom is Right
Ascension 0, increasing in RA in the anti-clockwise direction.
The possible \QU{} observing region is shown by solid lined box.} 
\label{fig:Kenfg}
\end{figure}

%The power spectra of the foreground models used are shown in Fig.
%\ref{fig:fgpower}. Here we see that for both EE and BB-spectra
%synchrotron emission dominates over vibrational dust, while the
%EE-spectra dominates over the foregrounds over the entire
%multipole range. For both EE and BB-spectra the 150 GHz channel
%will yield the lowest synchrotron foregrounds. Fig.
%\ref{fig:Kenfg} shows the polarized foreground levels expected in
%the region of sky accessible from the South Pole.

We closely follow T00 in
constructing the foreground power spectra required in the previous
section, and use the software provided on the associated website
\footnote{http://www.hep.upenn.edu/$\sim $max/foregrounds.html}.
\QU{} proposes to observe at frequencies of 100 and 150 GHz.
At these frequencies the relevant foregrounds are 
diffuse free-free emission, IR and radio point sources, 
synchrotron radiation and vibrating dust, rotating dust and thermal
Sunyaev-Zeldovich (SZ) radiation.
Each foreground is modelled using a spatial power spectrum,
$C_l\left(k\right)=\left(p\cal{A}\right)^2l^{-\beta}$, where $\beta$ gives the scale
dependence of the foreground fluctuations, $p$ is
the fraction polarized and $\cal{A}$ is the overall amplitude. A
frequency dependence is also defined and normalized to unity at a
reference frequency, $\nu_*$. For
point sources it is assumed that very bright sources (5~$ \sigma $ outlyers)
will be removed from the CMB maps, but that there will still be a residual
point source contamination after this subtraction. 
In T00, sets of estimates for these
parameters are given. We begin by using their
``middle-of-the-road'' foreground model. In this model, the 
only polarized foregrounds are sychrotron, dust and point sources. 
We then slightly modify this model to take into account recent observations
(Kovac et al. 2002; Bennett et al. 2003, hereafter B03).
These modifications lower the amplitude of the vibrating dust component and slightly
increase the amplitude of the synchrotron emission. The amplitudes then
roughly match those given in Fig.~10 of B03.  Also following B03
we have lowered the amplitude of the radio point sources and have neglected
rotating dust emission. The
values of those foreground parameters which are different from T00
are given in Table \ref{tb:fg}.

\begin{table}
 \begin{center}
  \caption{\label{tb:fg} Parameters used in foreground models which differ from
   those used in T00. All other parameters used are as per the
   ``middle-of-the-road'' model in T00.}
  \begin{tabular}{cccc}
   \hline
   Foreground  & Radio point & Synchrotron & Vibrating Dust \\
               & sources     &             &                \\
   \hline
   %$p$         & 0.0       & 0.13        & 0.0023           \\
   $\cal{A} /\mu \rm{K}$   & 0.66        & 95          & 7.5              \\
   $\nu_*$/GHz & -        & 20          & 90               \\
   \hline
  \end{tabular}
 \end{center}
\end{table}

The power spectra of the relevant foreground models
are shown in Fig. \ref{fig:fgpower} for the two \QU{} frequency bands.
The sychrotron radiation dominates the foregrounds at 
$ 100 $ GHz, whereas at 150 GHz both vibrating dust and 
sychroton radiation are important. The points sources only contribute
at very high multipoles.  For most of the multipole range of 
interest the EE spectrum dominates over the foregrounds. However,
for the smaller BB signal the total foreground contamination 
is larger than the signal of interest.

The analysis described in Section \ref{sec:fg1} gives the residual foreground
contamination given that foreground power spectra are well known or can be measured
from the experimental data. For \QU{} we assume that this is reasonable given
that other experiments, e.g. {\it WMAP} (B03), ARCHEOPS \cite{ARCH} and the recent Boomerang flight \cite{B2K},
will soon provide polarized maps at CMB frequencies. Recent advances in foreground
removal techniques \cite{Bacc:fg} indicate that it may be possible to remove some of the foreground noise 
from the signal. The residual foreground contamination used here therefore gives an upper limit
on that which can be expected in the final cleaned maps, given that our foreground models are accurate.

For a ground-based experiment it will also be
possible to select preferentially  regions of sky to observe
in which the foreground fluctuations are small, and so the foreground noise
can be reduced further. Fig. \ref{fig:Kenfg}
shows the region of sky which would be accessible to \QU{} from the South Pole
and the estimated levels of foreground contamination across this 
area from dust \cite{FD} and synchroton \cite{Gi} using the modified foreground models. A possible observing patch for \QU{} is also shown in which 
the mean foreground amplitude is low.  However, a more detailed
analysis will be performed to choose the final observing patch
with the lowest possible foreground
variance across the mutipole range of interest.
For these reasons we perform the relevant calculations
once using the full-sky foreground models descibed here, and again assuming
that the foregrounds are negigible.

We conclude that while our current understand of polarized
foregrounds is evolving, the expected level of foreground
contamination in the EE power spectrum should not be significant.
The GL component of the BB power spectrum should also be detectable
if patches of sky with low foreground variance can be targeted
or if foreground removal techniques can be successfully implemented.
This would also mean that the GW B-mode component should be 
measureable if the tensor-to-scalar ratio is large.
%\subsection{Constrained foregrounds}

%While all-sky surveys sample the full foreground distribution, a
%ground-based experiment can preferentially select out a region of
%sky with low foreground and carry out a long-term integration.
%This can have two purposes. The first is to lower the mean
%foreground in that region, so that the foreground does not swamp
%the intrinsic CMB polarization signal. The second effect is that
%pre-selection of the foreground constrains the power spectrum of
%the foregrounds. In effect the foreground modes must add up to
%give the known low mean foreground. If we assume the foreground
%can be modelled by a random Gaussian field, and we constrain the
%mean of the $Q$-field in a circular patch of radius $R$, the power
%spectrum of $E$-modes relative to the mean in that region is given
%by
% \be
%    \lgl C^{XY}_\ell| \overline{Q} \rgl = C^{XY}_\ell ( 1- 2 \pi
%        |W(\ell R)|^2)
% \end{equation}
% where
% \be
%    W(z) = \frac{2}{z^2} \left( 2 J_0(z) + z J_1(z) -2 \right)
%\end{equation}
% for a circular patch. We derive this result in Appendix B. We
% shall refer to such a low-foreground as a clean sky (CS) survey.

%% file: param.tex
In this Section we investigate the contribution which can be made
by ground-based polarization experiments to the measurement of the
cosmological parameters. Previous work on CMB parameter estimation
(Efstathiou \& Bond 1999; Zaldarriaga, Spergel \& Seljak 1997; Bond, Efstathiou \& Tegmark 1997) has
shown that the polarization data which can by obtained by the
forthcoming {\it WMAP} and {\it Planck} satellite missions will allow
a more accurate determination of many of the key cosmological
parameters. For a satellite experiment, this is mainly because the
degeneracy between $ \tau $ and $ A $ can be broken by measuring
the re-ionization bump in the polarization power spectra.  These
re-ionization bumps also create a high GW B-mode signal at low $
\ell $, so a full-sky measurement will also tighten constraints on
the tensor-to-scalar ratio.

\begin{figure*}
\hspace{0.5cm}
%\centering
\psfig{file=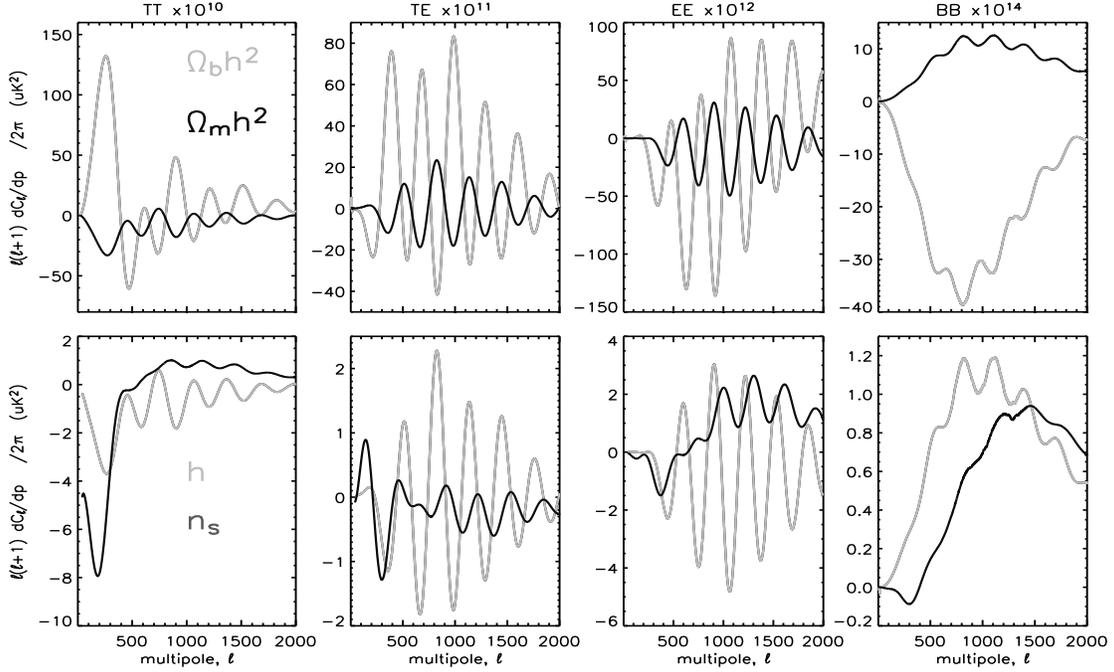,width=15.0cm,height=9.0cm}
%\epsfxsize=15cm \epsfysize=9cm
%\epsffile{fig8.ps} 
\caption{The derivative of
the CMB polarization power spectra with the parameters $\Omega_B
h^2$, $\Omega_mh^2$, $h$, and $n_s$. The models are generated by
CMBfast, with fiducial parameters given in the text.}
\label{fig:deriv1}
\end{figure*}

\begin{figure*}
%\centering
\hspace{0.5cm}
\psfig{file=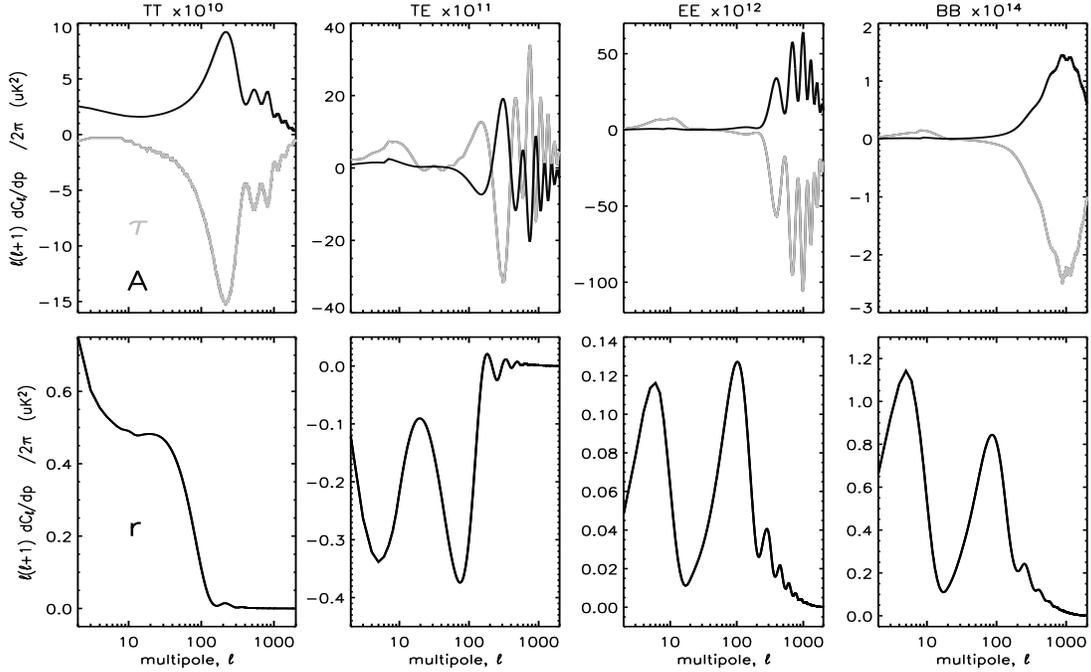,width=15.0cm,height=9.0cm}
%\epsfxsize=15cm
%\epsfysize=9cm
%\epsffile{fig9.ps}
\caption{The derivative of
the CMB polarization power spectra with the parameters $\tau$, $A$
and $r$. The models are generated by CMBfast, with fiducial
parameters given in the text. Note that the plot is now logarithmic
in order that the low $ \ell $ degeneracy breaking between $ \tau $ 
and $ A $ can be observed in the polarization power spectra.} \label{fig:deriv2}
\end{figure*}

\begin{figure*}
\hspace{4cm}
\psfig{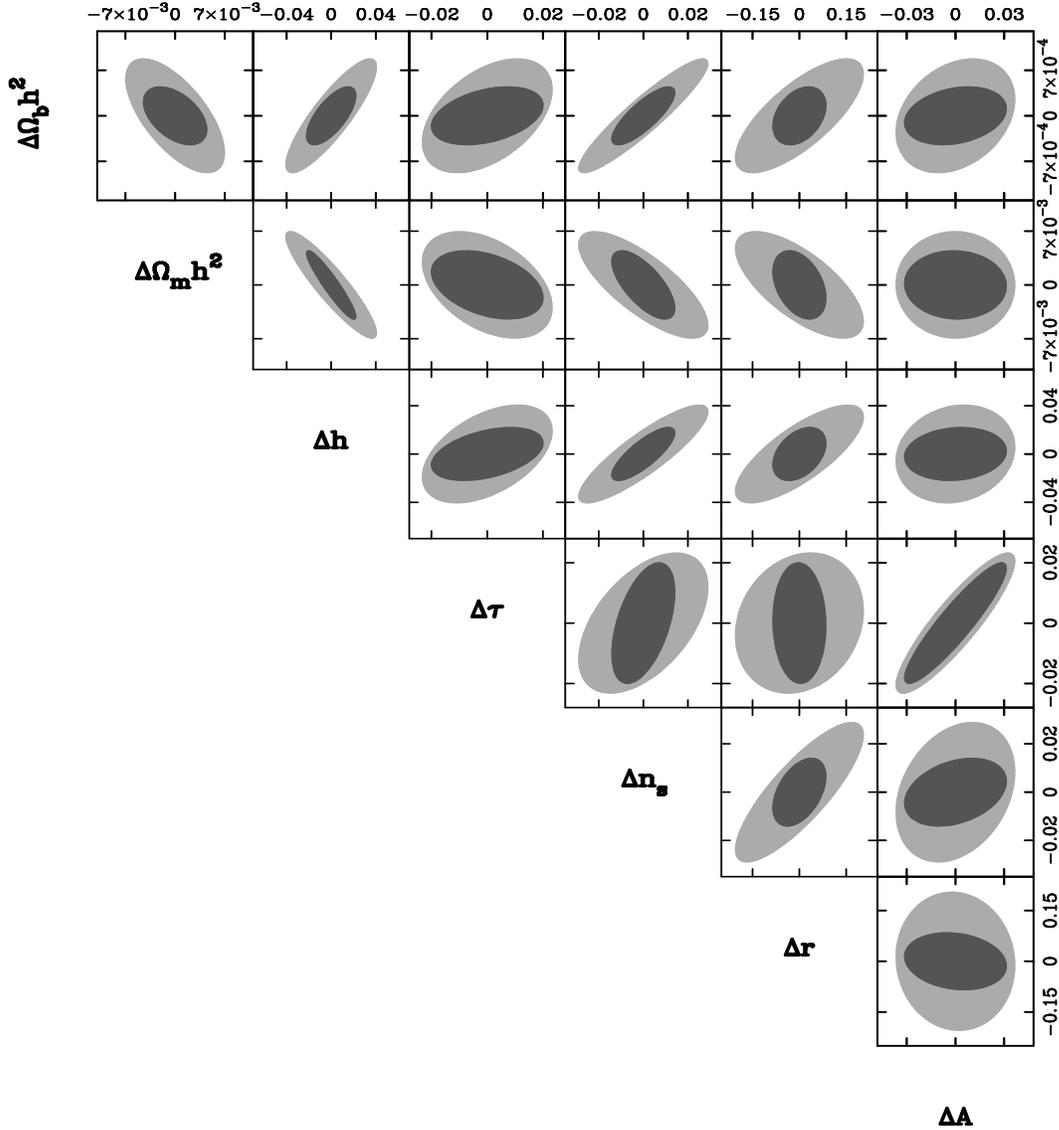}
%\epsfxsize=14cm
%\epsfysize=15cm
%\epsffile{fig10.ps}
\caption{Marginalized Fisher matrix relative parameter error
constraints ($\Delta \ln L=-1/2$)
anticipated for four-year {\it WMAP} results only (dark)
and four-year {\it WMAP} combined with \QU{} (light) for $r=0.01$ with foregrounds.
The projections of the ellipse onto the two axes give the standard errors
on each parameter. For a two-parameter $68$ per cent confidence region,
the ellipses should be scaled by a factor 1.5.}
\label{fig:ellipses}
\end{figure*}

As we have discussed, a ground-based polarization experiment can
concentrate on smaller areas of sky at higher resolution and so
can make a good measurement of the acoustic peaks out to high  $
\ell $ in the EE-power spectrum.  The information from a ground-based
experiment will therefore complement the full-sky
satellite data. It is also possible to choose an observing
strategy with targets the GW signal peak at intermediate scales
($\ell=100$). Ground-based constraints on the B-mode GW signal will
therefore also complement those obtainable from the current
generation of satellite experiments.

Finally, it is important to note that a CMB polarization
experiment is not just adding more data. A similar experiment
measuring only the temperature spectrum, over the same multipole
range, and with the same detector sensitivity, would add very
little new information as far as cosmological parameters are
concerned, although a high-$\ell$ temperature surveys may well
start to probe higher-order CMB effects. Hence polarization adds
unique information from the CMB.

We investigate the potential increase in the precision of the
measurement of cosmological parameters which can be achieved with a
ground-based experiment by comparing the expected four-year
results from {\it WMAP} alone to those which could be achieved by
combining \QU{} and {\it WMAP} data. To compare the two cases we
calculate the inverse Fisher matrix using equation
(\ref{eq:fishc}) to find the variances and covariances between
each of the parameters. For \QU{} we use the instrument model
dicussed in Section \ref{sec:QU}. The experimental parameters used for {\it WMAP} are given
in Table \ref{tb:sat}.

To calculate the derivatives in parameter space required in
equation (\ref{eq:fishc}) we use second-order differencing between
{\small CMBFAST} models for accuracy, with the corresponding parameter
changed up and down by 1 per cent. The derivative of the power spectrum
encapsulates the response of the spectrum to a change
in a particular parameter and hence quantifies its information
content. However, if the shape of the derivative for any two
parameters is too similar then the two parameters will be
degenerate and cannot both be constrained. The derivatives used in
the calculation are shown in Fig. \ref{fig:deriv1} and
\ref{fig:deriv2}. For most parameters, the shape of the
derivatives reflects the acoustic peaks in the power spectrum,
indicating that both information about parameters and their
differences are contained in the peaks. For instance with
temperature only $\Omega_m h^2$ and $\Omega_B h^2$ are quite
anti-correlated, but their derivatives oscillate out of phase for
TE and EE-spectra, breaking this degeneracy. Much of the
difference between $h$ and $n_s$ occurs in the low multipoles, but
there is a large difference at high-$\ell$ in the BB-spectra due
to the effects of gravitational lensing on these modes.

Fig. \ref{fig:deriv2} clearly shows the anti-correlation that
arises between $ \tau $ and $ A $ when only temperature
information is available. This degeneracy is seen to be broken on
large scales by the differences in the responses of the
polarization power spectra. However, going to high $\ell$ in the
polarization spectra, these parameters become strongly degenerate
again. Hence we can expect that a ground-based polarization
survey, which will have difficulty reaching the lower multipole
range, will not contribute much to lifting the $A-\tau$
degeneracy. Conversely, with only temperature information, $h$ and $n_s$ are 
strongly degenerate, with much of the difference in response coming 
at very low modes, or modes beyond a few hundred. However, adding polarization 
information, especially TE at around $\ell$ of 100, and lensed BB modes at
high $\ell$, breaks this degeneracy due to their different responses

Finally with only a temperature spectrum, the response to $ r $ is
limited to the first hundred multipoles. The $ TE $ and $
EE $ derivatives show that there is useful information about $ r $
on intermediate scales in the polarization spectra up to around $
\ell = 500 $, but for these multipoles the scalar EE and TE power spectra are very high
and so it will be difficult to extract this information from the signal.
However, the BB derivative also shows structure at higher multipoles and 
will provide information on $ r $ if the tensor signal is higher than the 
scalar lensing signal at the scales of interest, or if the lensing signal can be removed.

Having considered the responses of the power spectra to our
parameter set, we now turn to estimating parameter uncertainties
from satellite and ground-based surveys. To test the validity of
this procedure we have calculated the accuracy achievable with the
one-year {\it WMAP} data, and found that our results are on good
agreement with the one-year {\it WMAP} quoted parameter errors
\cite{WMAP:pvalue}.

%\begin{figure*}
%\epsfxsize=14cm \epsfysize=15cm \epsffile{FIGURES/plotfull_fs1.ps}
%\caption{Marginalized Fisher matrix relative parameter error
%constraints ($\Delta \ln L=-1/2$)
%anticipated for four-year {\it WMAP} results only (dark)
%and four-year {\it WMAP} combined with \QU{} (light) for $r=0.01$ with foregrounds.
%The projections of the ellipse onto the two axes give the standard errors
%on each parameter. For a two-parameter $68$ per cent confidence region,
%the ellipses should be scaled by a factor 1.5.}
%\label{fig:ellipses}
%\end{figure*}

Fig. \ref{fig:ellipses} shows the relative error
ellipses (defined by the $\Delta \ln L=-1/2$ contour)
expected from a 4-year {\it WMAP} experiment (darker ellipses)
and from a combined 2-year ground-based \QU{} and 4-year {\it WMAP}
experiment (lighter ellipses) for our fiducial 7-parameter model,
marginalizing over the other parameters. The projection of this contour gives
the marginalized one-parameter, $1-\sigma$ error for each parameter. For 
a two-parameter $68$ per cent confidence region, the ellispes should be scaled 
by a factor 1.5.  We assume that the
TE-cross spectrum can be estimated in the overlap region. Here we
see that a significant improvement of around a factor 2 is made on
most of the parameter set by adding in a ground-based polarization
survey, despite the significant difference in survey size. For most
parameters, this comes from the high-multipole information in the
EE-spectra, but there is also important information in the
BB-spectra, in particular for $r$, $h$ and $n_s$.

The 1-$\sigma$ marginalized parameter uncertainties for {\it WMAP}
and \QU{} + {\it WMAP} are shown in Table \ref{tb:perror}. 
By including \QU{} the precision with which the parameters
can be measured is improved by around a factor of two in most cases.
This increase in accuracy arises
from the extra information in the EE-spectra from modes
$\ell \ge 100 $, and from the strong BB-spectral dependence on
small scales for $n_s$ and $h$. Again, for a
temperature survey alone Fig. \ref{fig:deriv1} and
\ref{fig:deriv2} indicate there is no useful information at
high-multipoles.

It is interesting to look at how the information from the B-mode spectrum influences 
the parameter estimation. To examine this the same calcuation was made, but with the B-mode 
information removed from the Fisher matrix. For {\it WMAP}, this did not change the 
parameter estimates significantly, except for a slight increase
in the error on $r$ ($ \sim 10$ per cent). For {\it WMAP} most of the information on $r$
must therefore come from the TT, TE and EE spectra, 
and not from the weak upper limit on the B-mode spectrum. 
For \QU{} we find  a slight increase in the errors on $h$ and $n_s$ ($ \sim 20$ per cent)
due to the loss of the information contained in the 
B-mode lensing signal. However, the error on $r$ more than doubles if B-modes 
are not included. The B-mode information from \QU{} must therefore make
a significant contribution to the $r$ constraint, even though \QU{} cannot make a strong
detection of the GW B-mode signal.

\begin{figure}
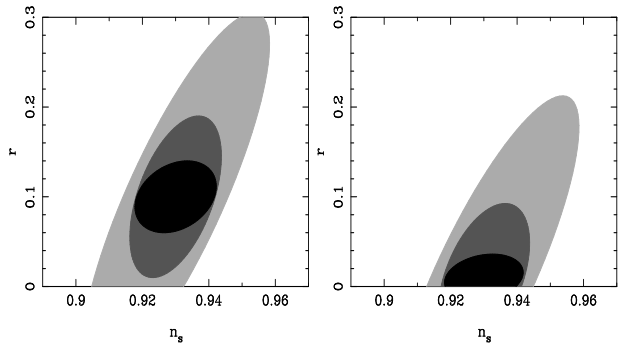

%\epsfxsize=4.0cm
%\epsfysize=4.5cm
%\subfigure{\epsffile{fig11a.ps}}
\subfigure{\psfig{file=fig11b.ps,width=4.0cm,height=4.5cm}}
%\epsfxsize=4.0cm
%\epsfysize=4.5cm
%\subfigure{\epsffile{fig11b.ps}}
\subfigure{\psfig{file=fig11a.ps,width=4.0cm,height=4.5cm}}
\caption{Predicted improvment on a joint measurement of $ r $ and $ n_s $ 
for two-year {\it WMAP} data (outer contour), two-year \QU{} + {\it WMAP} including
foregrounds (middle contour) and two-year \QU{} + {\it WMAP} without foregrounds (centre contour)
for $r=0.1$ (left) and $r=0.01$ (right). The $\Delta \ln L =-1/2$ contour is shown. For 68
per cent confidence limits, scale by a factor 1.5}
\label{fig:r_ns}
\end{figure}

Fig. \ref{fig:r_ns} shows the predicted improvement on a joint
measurement of $r$ and $n_s$ from a two-year \QU{} experiment and
four-year {\it WMAP} survey. With a detection of $r$, $n_s$ and the
amplitude $A$, the shape of the inflaton potential can be inferred \cite{HT}.

The poorest parameter improvement is for $ \tau $ and $ A $, which
only improve by  a factor of about $1.3 $. As discussed above
this is because the main differences appear on scales of 
$\ell \leq 100$, which are difficult to reach from the ground, but
accessible to satellite surveys.

We find that for most parameters the errors do not decrease significantly if the  
foreground contamination is completely removed. However, this is not the 
case for $ r $, where the error decreases by a factor of three if the foreground
contamination can be removed, leading to a factor of six improvement over 
the {\it WMAP}-only constraints. This can be clearly seen from the inner
contours in Fig. \ref{fig:r_ns}. This is because the foreground removal allows a much better
measurement of the B-mode GW signal to be made.  

\begin{table}
\caption{\label{tb:perror} Fisher matrix estimates of parameter errors}
\begin{tabular}{|c|c|c|c|}
\hline  Parameter & Value & {\it WMAP} & {\it WMAP} +\QU{} \\ \hline
 $ \Omega_bh^2 $  & 0.0224 & 0.0009 & 0.0004  \\
 $  \Omega_mh^2 $ & 0.135  & 0.007 & 0.004  \\
 $ h $            & 0.71   & 0.040 & 0.021  \\
 $ \tau $         & 0.17   & 0.023 & 0.020  \\
 $ n_s $          & 0.93   & 0.029 & 0.014  \\
 $ r $            & 0.01 (0.1) & 0.206 (0.203) & 0.082 (0.090) \\
 $ A $            & 0.83   & 0.036 & 0.031  \\
\hline
\end{tabular}
\end{table}

%% file: appendix.tex
There are a number of definitions for the sensitivity of a CMB
polarization experiment and this is often a cause of confusion
when comparing different sensitivity parameters.  For a total-
power CMB experiment the sensitivity is usually defined in terms
of the noise equivalent temperature (NET). This is the signal
needed from the source to give a signal-to-noise ratio of unity in
a one-second integration time.\footnote{For CMB work the
sensitivity is usually quoted as an NET, in units of $ Ks^{1/2}
$, instead of as a noise equivalent power (NEP), which is normally
used in sub-millimetre astronomy.  This makes it easier to combine
experimental work with theory, as the power spectra ($ C_{\ell} $)
are defined in terms of temperature units. The NEP is normally
quoted per unit bandwidth and so has units of $ WHz^{-1/2}  $,
which is equivalent to noise produced in a  half second
integration time.  To change NET in $ Ks^{1/2} $ to NEP in $
WHz^{-1/2}  $ the conversion is:
\begin{equation}
NET(Ks^{1/2})=\frac{NEP(WHz^{-1/2})}{\sqrt2 \:\:  \partial
B_{\nu}/\partial T},
\end{equation}
where $ \partial B_{\nu}/\partial T $ is the derivative of  the
source (the CMB) with respect to temperature. The factor of $
\sqrt{2} $ converts from $ Hz $ to seconds.} To measure
polarization an equivalent definition is required in terms of the
Q and U Stokes parameters. For a linearly polarized source of
total intensity, $ I $, of which a fraction $ p $ is polarized at
an angle $ \chi $ to the reference direction, the Stokes
parameters can be defined as:
\begin{eqnarray}
Q=pI\cos(2\chi), \nonumber \\
U=pI\sin(2\chi).
\end{eqnarray}
If we orientate the axis of the reference system so that it is
aligned with the polarization angle of the source ($ \chi=0 $)
then we have $ U=0 $ and $ Q=pI $ so that $ Q $ gives the total
polarized intensity. We can then define the polarization
sensitivity, NEQ, as the polarized signal from the source needed
to give a signal-to-noise ratio of unity in a one-second
integration time for a source with a polarization angle aligned
with the reference direction of the measurement.

For \QU, the polarized measurements will be made with pairs of
polarization-sensitive bolometers (PSBs). The two bolometers in a
PSB pair are sensitive to orthogonal polarization states of the
incoming radiation. The intensity measured by the co-polar ($x$)
and cross-polar ($y$) device is given by \cite{PSBs}:
\begin{eqnarray}
I_x=\frac{1}{2}(I+Q),  \nonumber \\
I_y=\frac{1}{2}(I-Q),
\end{eqnarray}
in a reference system aligned with the polarization angle  of
the source.  The total intensity is found by adding the two
bolometer outputs and the $Q$ Stokes parameter is found by
differencing the outputs.

For a bolometer, the noise equivalent power due to photon noise
(NEP) is given by \cite{Lamarre}:
\begin{equation}
\label{noise}
NEP^2=2h\nu P+\frac{2P^2}{m \Delta \nu}
\end{equation}
where is $m$ is the number of polarization states detected  ($m$
is either 1 or 2). For a single PSB, $ m=1 $,  as only a single
polarization state is detected. $ P $ is the power in a band of
width $ \Delta \nu $:
\begin{equation}
P_{\nu}=\eta_d \eta_t \varepsilon A\Omega B_{\nu}\Delta \nu,
\end{equation}
where $ A\Omega $ is the throughput of the system, $ \varepsilon $
is the emissivity of the source and $ B_{\nu} $ is the intensity
of the radiation that would be emitted from a perfect black
body. The total efficiency of the system is $\eta=\eta_{d}\eta_{t}$, where $\eta_{d}$ is the 
detector efficiency and $\eta_{t}$ is the instrument efficiency. We assume that a lossless
PSB will absorb half of the incident unpolarized radiation, giving
$ \eta_d=1/2 $.  The NET due to photon noise in each PSB from the
unpolarized background radiation is therefore:
\begin{equation}
\label{eq:single} NET_s=\frac{(2h\nu P +2P^2/\Delta \nu)^{1/2}}{
\eta_d \eta_t \:\: \partial B_{\nu}/\partial T },
\end{equation}
where the factor of $ \eta_d \eta_t  $ is needed to convert from the noise
at the detector to the signal required at the source, as only a fraction of the radation from
the source will be absorbed by the PSB.
 $  \partial B_{\nu}/\partial T  $ is the derivative of the source
intensity with respect to temperature and converts from an NEP to
an NET. Equation (\ref{eq:single}) gives the NET for a measurement
of the {\it temperature} of the CMB with a single PSB.

In order to measure the polarization we require a pair of PSBs.
The temperature sensitivity of a PSB pair can be obtained by {\it averaging} the two
outputs so that:
\begin{equation}
NET_{pair}=NET_s/\sqrt{2}.
\end{equation}
$ NET_{pair} $ is exactly the NET that would be obtained if a single
normal (not polarization sensitive) bolometer had been used. For a
measurement of Q, the two outputs are {\it differenced} so that:
\begin{equation}
NEQ=\sqrt{2} \eta_d NET_s=\frac{NET_s}{\sqrt{2}}.
\end{equation}
An important point to note is the factor of $ \eta_d $ in this
expression.  This is because we are now measuring the
signal from a polarized source, so the factor of $ \eta_d $
which was needed in equation \ref{eq:single} to find the noise for a measurement of the
total power is no-longer required. All of the polarized
radiation is absorbed by a single PSB when it is correctly aligned
with the polarization angle of the source.

When defining the sensitivity of a PSB it is therefore important
to state whether a sensitivity is an NET for a single detector, an
NET for a pair of detectors, or an NEQ for a pair of detectors. The
expression for the pixel noise given in Section \ref{sec:formal}
will depend on the sensitivity definition used:
\begin{equation}
\sigma^2=\frac{NET^2 \Theta^2}{t_{obs}N_{PSB}\Omega_{pix}^2}.
\end{equation}
If the NET is for a single PSB (as in Table \ref{tb:inst} ), $ N_{PSB} $ is the total number of PSBs.  If the sensitivity is given as an NEQ for a PSB pair, $ N_{PSB} $
is the number of pairs.

%% file: CMBpol_gbase.bbl
\begin{thebibliography}{}
\bibitem[\protect\citename{Baccigalupi} 2003]{Bacc:fg}
Baccigalupi C., 2003, New Astron. Rev., 47, 1127 

\bibitem[\protect\citename{Bennett et al.} 2003]{WMAP:fg}
Bennett C. L. et al., 2003, ApJ, 148, 97

\bibitem[\protect\citename{Benoit et al.} 2003]{ARCH}
Benoit A. et al., 2003, A\&A, 339, L19

\bibitem[\protect\citename{Bond, Efstathiou \& Tegmark} 1997]{BET}
Bond J. R., Efstathiou G., Tegmark M., 1997, MNRAS, 291, L33-L41

\bibitem[\protect\citename{Bunn} 2002]{Bunn}
Bunn E. F., 2002, Phys.Rev.D, 65, 043003

\bibitem[\protect\citename{Church et al.} 2003]{SEC}
Church S. E., et al., 2003, New Astron. Rev., 47, 1083

\bibitem[\protect\citename{Efstathiou \& Bond} 1999]{EB}
Efstathiou G., Bond J. R., 1999, MNRAS, 304, 75

\bibitem[\protect\citename{Finkbeiner, Davis \& Schlegel} 1999]{FD}
Finkbeiner D. P., Davis M., Schlegel D. J., 1999, ApJ, 524, 867  

\bibitem[\protect\citename{Giardino et al.} 2002]{Gi}
Giardino G., Banday A. J., Górski K. M., Bennett K., Jonas J. L., Tauber J.
2002, A\&A, 387, 82

\bibitem[\protect\citename{Guzik et al.} 2000]{Guzik}
Guzik J., Seljak U., Zaldarriaga M., 2000, Phys.Rev.D, 64, 043517

\bibitem[\protect\citename{Lay \& Halverson} 1998]{atm}
Lay O. P., Halverson, N. W., 2000, ApJ, 543, 787

\bibitem[\protect\citename{Hobson et al.} 1998]{fgremove}
Hobson M. P., Jones A. W., Lasenby A. N., Bouchet F. R.
1998, MNRAS, 300, 1, 29

\bibitem[\protect\citename{Hobson \& Magueijo} 1996]{Hobson}
Hobson M. P.,  Magueijo J., 1996, MNRAS, 283, 4, 1133

\bibitem[\protect\citename{Hoffman \& Turner} 2001]{HT}
Hoffman M. B., Turner M. S., 2001, Phys.Rev.D, 64, 2, 023506

\bibitem[\protect\citename{Hu} 2001]{HuFish}
Hu W., 2001, Phys.Rev.D, 65, 023003

\bibitem[\protect\citename{Hu \& White} 1997]{Hreview}
Hu W., White M., New Astron., 2, 323

\bibitem[\protect\citename{Jaffe et al.} 2000]{JKW}
Jaffe A. H., Kamionkowski M., Wang L., 2000, Phys.Rev.D, 61, 083501

\bibitem[\protect\citename{Jones et al.} 2003]{PSBs}
Jones W. C., Bhatia R., Bock J. J., Lange A. E., 2003, in Phillips, T. G., Zmuidzinas, J., eds,
Proc. SPIE, 4855, 227

\bibitem[\protect\citename{Kamionkowski et al.} 1997]{KKS}
Kamionkowski M., Kosowsky A., Stebbins A., 1997, Phys.Rev.D, 55, 7368

\bibitem[\protect\citename{Kaplinghat, Knox \& Song} 2003]{neutrino}
Kaplinghat M., Knox K., Song Y., 2003, Phys. Rev. Lett., 91, 24301

\bibitem[\protect\citename{Keating et al.} 1998]{BK}
Keating B., Timbie P., Polnarev A., Steinberger  J., 1998, ApJ, 495, 580

\bibitem[\protect\citename{} 2002]{Kesden}
Kesden M., Cooray A., Kamionkowski M., 2002, Phys.Rev.Lett, 89, 1, 011304

\bibitem[\protect\citename{Knox \& Song 2002, Kesden, Cooray \& Kamionkowski} 2002]{KS}
Knox L., Song S., 2002, Phys.Rev.Lett, 89, 1, 011303

\bibitem[\protect\citename{Kogut et al.} 2003]{WMAP:TE}
Kogut A. et al, 2003, ApJ, Suppl., 148, 161

\bibitem[\protect\citename{Kovac et al.} 2002]{DASI}
Kovac J., Leitch E. M., Pryke C., Carlstrom J. E, Halverson N. W., Holzapfel W. L.,
2002, Nat, 420, 772

\bibitem[\protect\citename{Lamarre} 1986]{Lamarre}
Lamarre  J., 1986, Applied Opt., 25, 870

\bibitem[\protect\citename{Leach \& Liddle} 2003]{LL}
Leach S. M., Liddle A. R., 2003, Phys.Rev.D, 68, 123508

\bibitem[\protect\citename{Lewis et al.} 2002]{Lewis}
Lewis A., Challinor A., Turok N., 2002, Phys.Rev.D, 65, 2, 023505

\bibitem[\protect\citename{Maino et al.} 2002]{Maino}
Maino D., et al., 2002, MNRAS, 334, 1, 53

\bibitem[\protect\citename{Montroy et al.} 2003]{B2K}
Montroy T. et al., 2003, New Astron. Rev., 47, 1057

%\bibitem[\protect\citename{Peiris et al.} 2003]{WMAP:infl}
%Peiris H. V., et al., 2003, ApJ, accepted, (astro-ph/0302225)

%\bibitem[\protect\citename{Runyan et al.} 2003]{ACBAR}
%Runyan et al., 2003, ApJ, submitted, (astro-ph/0303515)

%\bibitem[\protect\citename{Rusholme et al.} 2003]{Ben}
%Rusholme B. A., 2003, Carnegie Observatories Astrophysics Series, Vol. 2: Measuring and Modeling the Universe, %ed. W. L. Freedman, Pasadena: Carnegie Observatories, \\http://www.ociw.edu/ociw/symposia/series/symposium2/ 

\bibitem[\protect\citename{Seljak \& Zaldarriaga} 1996]{SZ}
Seljak U., Zaldarriaga M., 1996, ApJ, 469, 473

\bibitem[\protect\citename{Spergel et al.} 2003]{WMAP:pvalue}
Spergel D. N., et al., 2003, ApJ Suppl., 148, 175

\bibitem[\protect\citename{Tegmark el al.} 2000]{TEHO}
Tegmark M., Eisenstein D. J., Hu W., Oliveira-Costa A., 2000, ApJ, 530, 133

\bibitem[\protect\citename{Tegmark, Taylor \& Heavens} 1997]{Andy}
Tegmark M., Taylor A. N., Heavens A. F., 1997,  ApJ,  480,  22

\bibitem[\protect\citename{Turner \& White} 1996]{TW}
Turner, M., S., 1996, Phys.Rev.D, 53, 6822

\bibitem[\protect\citename{Verde et al.} 2003]{WMAP:pmethod}
Verde L., et al., 2003, ApJ Suppl., 148, 195

\bibitem[\protect\citename{Zaldarriaga} 2003]{Zreview}
Zaldarriaga, M., 2003, American Astronomical Society Meeting, 202, 5601Z

\bibitem[\protect\citename{Zaldarriaga \& Seljak} 1997]{ZS}
Zaldarriaga, M., Seljak, U., 1997, Phys.Rev.D, 55, 1830

\bibitem[\protect\citename{Zaldarriaga et al} 1997]{ZSSp}
Zaldarriaga, M., Spergel, D. N., Seljak, U., 1997, ApJ, 488, 1

%\bibitem[\protect\citename{} ]{}
%\bibitem[\protect\citename{} ]{}
%\bibitem[\protect\citename{} ]{}
%\bibitem[\protect\citename{} ]{}
\end{thebibliography}
